\documentclass[aps,prc,floatfix,showpacs,twocolumn]{revtex4-1}
\usepackage{ulem}
\usepackage{dcolumn}
\usepackage{bm}
\usepackage{color}
\usepackage{amssymb}
\usepackage{amsmath}
\usepackage{graphicx}
\usepackage{amsfonts}
\usepackage{slashed}
\usepackage{pstricks}
\usepackage{float}
\usepackage{hyperref}
\usepackage{array}
\allowdisplaybreaks
\usepackage{dcolumn}
\usepackage{epsf}
\begin{document}
\title{Relativistic effects in ab-initio electron-nucleus scattering}
\author{
Noemi Rocco$^{\, {\rm a} }$,
Winfried Leidemann$^{\, {\rm b,c} }$,
Alessandro Lovato$^{\, {\rm c,d} }$,
Giuseppina Orlandini$^{\, {\rm b,c} }$
}
\affiliation{
$^{\,{\rm a}}$\mbox{Department of Physics, University of Surrey, Guildford, GU2 7HX, UK}\\
$^{\,{\rm b}}$\mbox{Dipartimento di Fisica, Universit\'a di Trento, Via Sommarive 14, I-38123 Trento, Italy}\\
$^{\,{\rm c}}$\mbox{INFN-TIFPA Trento Institute of Fundamental Physics and Applications, Via Sommarive, 14, 38123 Trento, Italy}\\
$^{\,{\rm d}}$\mbox{Physics Division, Argonne National Laboratory, Argonne, Illinois 60439, USA}\\
}
\date{\today}

%
\date{\today}
\begin{abstract} {
The electromagnetic responses obtained from Green's 
function Monte Carlo (GFMC) calculations are based on realistic treatments of nuclear interactions and currents. The main 
limitations of this method comes from its nonrelativistic nature and its computational cost, the latter hampering the direct
evaluation of the inclusive cross sections as measured by experiments. We extend the applicability of GFMC in the quasielastic
region to intermediate momentum transfers by performing the calculations in a reference frame that minimizes nucleon
momenta. Additional relativistic effects in the kinematics are accounted for employing the two-fragment model. 
In addition, we developed a novel algorithm, based on the concept of first-kind scaling, to compute the inclusive electromagnetic
cross section of $^4$He through an accurate and reliable interpolation of the response functions. A very good agreement is 
obtained between theoretical and experimental cross sections for a variety of kinematical setups. 
This offers a promising prospect for the data analysis of 
neutrino-oscillation experiments that requires an accurate description of nuclear dynamics in which 
relativistic effects are fully accounted for.}
\end{abstract}
\pacs{24.10.Cn,25.30.Pt,26.60.-c}
\maketitle

\section{Introduction}
The analysis of neutrino-nucleus interactions in the broad kinematic region relevant for 
the present~\cite{minerva_web,mb_web,nova_web,t2k_web} 
and future~\cite{dune_web,hk_web} generation of neutrino oscillation experiments requires an 
accurate understanding of nuclear dynamics. 
The relevance of nuclear models is critical to the reconstruction of the initial neutrino energy, 
even in experiments where both near and far 
detectors are present~\cite{Katori:2016yel}.
Recent experimental studies of neutrino-nucleus interactions have provided ample evidence of the inadequacy of the relativistic
 Fermi gas model, routinely employed in event generators, to describe the observed cross section. The complexity of nuclear 
 dynamics and the variety of reaction mechanisms are such that \textit{ab-initio} calculations of nuclear structure and electroweak 
 interactions with nuclei are necessary~\cite{Benhar:2015wva}.

Within nuclear \textit{ab-initio} approaches the nucleus is treated as an assembly of nucleons interacting with 
each other via two- and three-body effective 
potentials~\cite{Lei13,Barrett:2013nh,Hagen:2013nca,Hergert:2015awm,Carbone:2013eqa,Epelbaum:2011md,Carlson:2014vla}. 
The interaction with external electroweak probes is described by one- and two-body effective currents that are consistent 
with the nuclear interaction. Hence, properties of few-body nuclear systems, such as the nucleon-nucleon (NN)
scattering data and the
 binding energies of light nuclei, ultimately constrain the current operators~\cite{Hammer:2012id}. 
 This is particularly apparent for the electromagnetic
 longitudinal  current, which is connected to the nuclear potential through the continuity equation. 

The Green's Function Monte Carlo (GFMC) approach is an ab-initio method that allows for a very accurate description
of the structure and low-energy transitions of $A\leq 12$ nuclei~\cite{Carlson:1987zz}. More recently, exploiting integral transform techniques, the
 GFMC method has also been applied to the calculation of the  electromagnetic
response functions of $^4$He and $^{12}$C, giving a full account of the dynamics of the constituent 
nucleons in the quasielastic sector. Once two-body currents are accounted for, 
GFMC predictions are in very good agreement with experimental data~\cite{Lovato:2015,Lovato:2016gkq}. 
However, considering that
explicit pion degrees of freedom are not taken into account the strength seems to be somewhat too large beyond pion threshold.

As a matter of fact, at higher momentum transfer the applicability of GFMC to electroweak scattering and in 
particular to the analysis of neutrino-nucleus scattering is hampered by its nonrelativistic 
nature. Whilst leading relativistic corrections are included in the current operators, 
the quantum mechanical framework is nonrelativistic. The strategy introduced
in Ref.~\cite{Amaro:2005dn} to account for relativistic  kinematics in nonrelativistic calculations 
can only be reliably applied to independent particle models of nuclear dynamics.

The inclusion of relativistic corrections in a more sophisticated approach 
has been 
first discussed in Ref.~\cite{Efros_LOT:2005}.
The Authors argued that performing the nonrelativistic calculation in a specific reference frame can minimize the error introduced by the 
approximate treatment of relativistic effects. In addition, the frame dependence of nonrelativistic results 
can be reduced using the so-called 
``two-fragment model'' to obtain, in a relativistically correct way, the kinematic inputs 
of the nonrelativistic dynamical calculation
. This approach has been successfully employed in the \textit{ab-initio} calculation of the 
electromagnetic longitudinal~\cite{Efros_LOT:2005} and transverse~\cite{Efros_LOT:2010,Yuan:2010gh,Efros_LOT:2011,Yuan:2011rd} 
response functions of $^3$He at intermediate
momentum transfers (up to $|{\bf q}|$=700 MeV).

Following Ref.~\cite{Efros_LOT:2005}, in this work we gauge the role of relativistic effects in the original GFMC 
electromagnetic response functions of 
$^4$He by studying their frame dependence with and without the two-fragment model.

The GFMC calculations of the response functions
and cross sections by neutral-current scattering of neutrinos off $^{12}$C have been recently presented in Ref.~\cite{Lovato:2017cux}.
In that work, the neutral current differential cross sections have been computed for a single value of the momentum transfer, 
$|{\bf q}|=$ 570 MeV, as a function of the energy loss $\omega$. On the other hand, experimental electron- and neutrino-nucleus 
cross sections are commonly given for fixed values of the incoming beam energy and scattering angle. Their direct evaluation requires
to perform GFMC calculations of the nuclear response functions for several values of $|{\bf q}|$, whose computational cost exceeds 
the current availability. In this work we developed a novel algorithm suitable to compute the double differential cross sections of
electron-$^4$He scattering through an efficient interpolation of the available nuclear responses. The latter exploits the scaling features
of the GFMC electromagnetic response functions, which have been recently investigated in Ref.~\cite{Rocco:2017hmh}.
Using this algorithm and employing the relativistic treatment mentioned above we perform an extensive comparison
of our results with the  electron scattering data, 
for initial electron energies ranging from 0.3 GeV to 1.1 GeV.

In Section~\ref{sec:def:meth} we shortly review the formalism connecting  
the electron-nucleus cross section to the longitudinal and transverse 
response functions and discuss the main elements of their calculation within the GFMC approach. In addition we make 
a comparison to results obtained with the Lorentz Integral Transform (LIT) method~\cite{Efros_LO:1994,Efros:2007nq}.
In Section~\ref{sec:rel:eff} we review the approach of Ref.~\cite{Efros_LOT:2005} to account for relativistic effects and  
study the frame dependence of the GFMC responses, as well as its reduction with the two-fragment model. 
Section~\ref{sec:cross:sec} is devoted to the calculation of the electron-$^4$He differential cross sections and 
to the comparison with experiment. 
Finally, in Section~\ref{sec:concl}
we draw our conclusions.

\section{FORMALISM} 
\label{sec:def:meth}
In the one-photon-exchange approximation, the inclusive double differential electron-nucleus cross section can be written 
in terms of the two response functions, $R_L({\bf q}, \omega)$ 
and $R_T({\bf q},\omega)$, describing interactions with longitudinally (L) and transversely (T) polarized virtual photons
\begin{align}
\frac{d^2\sigma}{d E_{e^\prime} d\Omega_{e}} &  =\left( \frac{d \sigma}{d\Omega_{e}} \right)_{\rm{M}} \Big[  A_L\,  R_L(|{\bf q}|,\omega) \nonumber \\
& + A_T\,  R_T(|{\bf q}|,\omega) \Big] \ ,
\label{eq:x:sec}
\end{align}
where 
\begin{align}
A_L = \Big( \frac{q^2}{{\bf q}^2}\Big)^2  \ \ \ , \ \ \ A_T = -\frac{1}{2}\frac{q^2}{{\bf q}^2}+\tan^2\frac{\theta_e}{2}  \ , 
\end{align}
and
\begin{align}
\label{Mott}
\left( \frac{d \sigma}{d \Omega_{e}} \right)_{\rm{M}}= \left[ \frac{\alpha \cos(\theta_{e}/2)}{2 E_{e^\prime}\sin^2(\theta_{e}/2) }\right]^2
\end{align} 
is the Mott cross section. In the above equation $\alpha\simeq1/137$ is the fine structure constant, $E_e^\prime$ and $\theta_e$ are
the final lepton energy and scattering angle, respectively, {\bf q} and $\omega$ are energy and  momentum transferred 
by the electron to the target nucleus, and $q^2=\omega^2-{\bf q}^2$. 

The longitudinal and transverse response functions are expressed in terms of the nuclear current matrix elements
\begin{align}
R_{\alpha}(|{\bf q}|,\omega)&=\sum_f \langle 0| j_{\alpha}^\dagger(\mathbf{q},\omega) |f\rangle \langle f | j_{\alpha}(\mathbf{q},\omega) | 0\rangle  \nonumber\\
&\times \delta(\omega-E_f + E_0)
\label{eq:RLT}
\end{align}
where $|0\rangle$ and $|f\rangle$ represent the nuclear initial ground-state and final bound- or scattering-state of energies $E_0$
and $E_f$,
and $j_{\alpha}(\mathbf{q},\omega)$ ($\alpha=L,T$) denotes the longitudinal and transverse 
components of the electromagnetic current.
For moderate momentum transfer, corresponding to $|\mathbf{q}|\lesssim 500$ MeV, nonrelativistic nuclear many-body theory can be applied to
consistently describe the initial and the final scattering states in the quasielastic peak region. To this aim, a nonrelativistic reduction of the 
electromagnetic currents, which includes one- and two-body terms consistent with the nuclear Hamiltonian, is performed. The explicit 
expressions for the electromagnetic currents employed in this work can be found in Ref.~\cite{Shen:2012}

\subsection{The GFMC approach to Response Functions}

Following the strategy adopted in Refs.~\cite{Carlson:2002,Lovato:2015,Lovato:2016gkq}, instead of attempting 
a direct calculation of each individual 
transition amplitude $|0\rangle \to | f\rangle$, we exploit integral transform techniques to reduce the problem to a ground-state one. 
In particular, 
we evaluate the inelastic Euclidean responses, defined through the following Laplace transform of the electromagnetic response 
functions
\begin{equation}
{E}_\alpha(|\mathbf{q}|,\tau)=\int_{\omega_{el}^+}^\infty d\omega R_{\alpha}(|{\bf q}|,\omega) e^{-\omega\tau}\, ,
\end{equation}
where $\omega_{el}$ is the energy of the recoiling ground state. Besides the energy-conserving $\delta$-function, 
the response functions depend upon $\omega$ through the electromagnetic form factors of the nucleon and 
$N$-to-$\Delta$ transition in the currents. We artificially remove these dependences by evaluating the form factors at 
the quasielastic peak $q^2_{qe} = \omega^2_{qe}-{\bf q}^2$.
Exploiting the completeness of the final states of Eq.~\eqref{eq:RLT}, the inelastic Euclidean responses can be written as the 
following ground-state expectation value
\begin{align}
{ E}_{\alpha}(|\mathbf{q}|,\tau)\!&=\!\langle \Psi_0| j^\dagger_\alpha({\bf q},\omega_{qe}) e^{-(H-E_0)\tau} 
j_\alpha({\bf q},\omega_{qe}) |\Psi_0\rangle\nonumber\\
&-|F_{\alpha}(q)|^2 {\rm e}^{-\tau \omega_{\rm el}} \,
\label{eq:euc_me}
\end{align}
where $F_{\alpha}(q)$ is the longitudinal elastic form factor and $H$ the nuclear Hamiltonian. For its potential part we
use the Argonne $v_{18}$ (AV18)~\cite{Wiringa:1995} NN potential and Illinois-7 (IL7)~\cite{Pieper:2001} 
three-nucleon force (3NF). The Simon~\cite{Simon:1980hu}, Galster~\cite{Galster:1971kv}, and H\"ohler~\cite{Hohler:1976ax} parametrizations are used for the 
proton electric, neutron electric, and proton and neutron magnetic form factors, respectively.

In order to reduce the computational cost and to evaluate the terms in the currents
that depend upon the momentum of the nucleon, we use our best variational trial wave function
$|\Psi_T\rangle$ for $|\Psi_0\rangle$.  Hence, the response functions are those obtained from
$|\Psi_T\rangle$ rather than those from the evolved GFMC wave function. However, the sum rule results 
of Ref.~\cite{Lovato:2013} indicate that this is indeed a good approximation.

The calculation of this ground-state expectation value is carried out in two steps. At first 
the unconstrained imaginary-time propagation of $|\Psi_T\rangle$ is performed 
and stored. Then, the states obtained from $j_\alpha({\bf q},\omega_{qe}) |\Psi_T\rangle$ are
propagated in imaginary-time and the scalar product of $e^{-(H-E_0)\tau} 
j_\alpha({\bf q},\omega_{qe}) |\Psi_T\rangle$ with $\langle \Psi_T| j^\dagger_\alpha({\bf q},\omega_{qe})$
is performed on a grid of $\tau_i$ values (for more details see Refs.~\cite{Carlson:2002,Lovato:2015,Carlson:2015}).

The inversion of the Laplace transform, needed to retrieve the response functions, is performed 
exploiting maximum entropy techniques, as described in Ref.~\cite{Lovato:2016gkq}.
\subsection{Comparison with Lorentz Integral Transform results}\label{IIB}

\begin{figure}[H]
\centering
\includegraphics[scale=0.675]{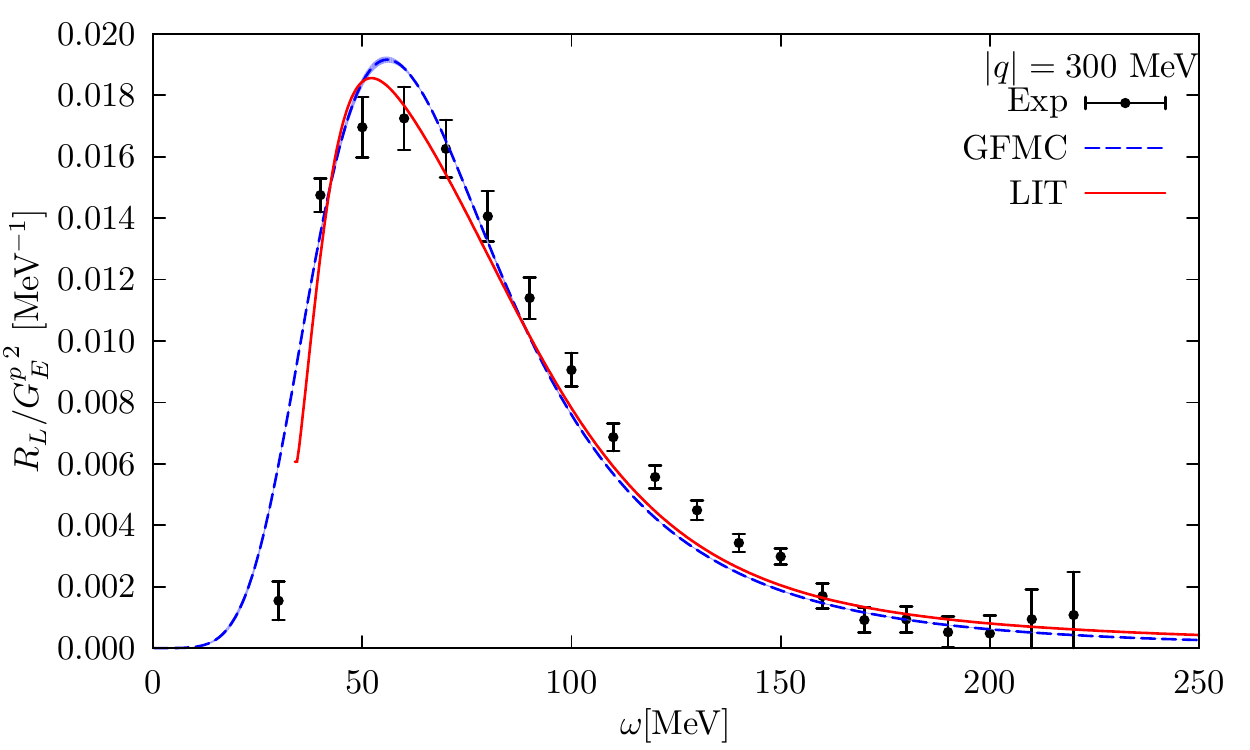}
\caption{ Longitudinal electromagnetic response functions of $^4$He at $|{\bf q}|=300$ MeV obtained inverting the 
Laplace and Lorentz integral transforms compared to the experimental data of Ref. \cite{Carlson:2002}.}
\label{fig:lit_laplace}
\end{figure}

To test the reliability of the GFMC calculation and in particular of the inversion procedure, in Fig.~\ref{fig:lit_laplace} we 
compare the longitudinal response function of $^4$He divided by the proton electric form factor 
squared with that obtained in Ref.~\cite{Bacca:2009,Bacca:2009prc}, employing the 
LIT method. The latter has been computed
representing $|0\rangle$  
and the LIT states $\tilde \Psi$ (see Ref.~\cite{Efros_LO:1994}) 
in terms of hyperspherical harmonics. The Hamiltonian used in that case was 
the NN AV18 potential and the Urbana IX (UIX) 3NF. The agreement with experimental data, taken from 
Ref.~\cite{Carlson:2002} is remarkably good. The two theoretical curves are also in very good
agreement. The small discrepancies can be ascribed to: i) the different 3NF models employed,
ii) the very narrow isoscalar monopole resonance contribution {(see ~\cite{Bacca:2012xv})} that has been subtracted 
from the LIT, iii) the spin-orbit correction in the longitudinal current operator that is only included 
in the GFMC results, iv) the use of a variational Monte Carlo ground state. Finally, it has to be noted that 
resolving the low-energy transfer region of the 
response requires imaginary-time evolution to large values of $\tau$, which is hampered by the Fermion sign problem.
At $|{\bf q}|=500$ MeV the difference between LIT and GFMC results becomes somewhat more pronounced. 
It mainly consists in a slightly shifted quasielastic peak position. We checked that the origin of the difference
is not due to an inversion problem. In fact, in addition to the standard LIT inversion method~\cite{Barnea:2009zu}, we used the 
maximum entropy technique to invert the LIT. We did not find significant differences in the resulting $R_L$. 
It remains object of further future investigations whether the differences can be explained by the items i)-iv)
mentioned above.
\begin{figure}[h]
\centering
\includegraphics[scale=0.675]{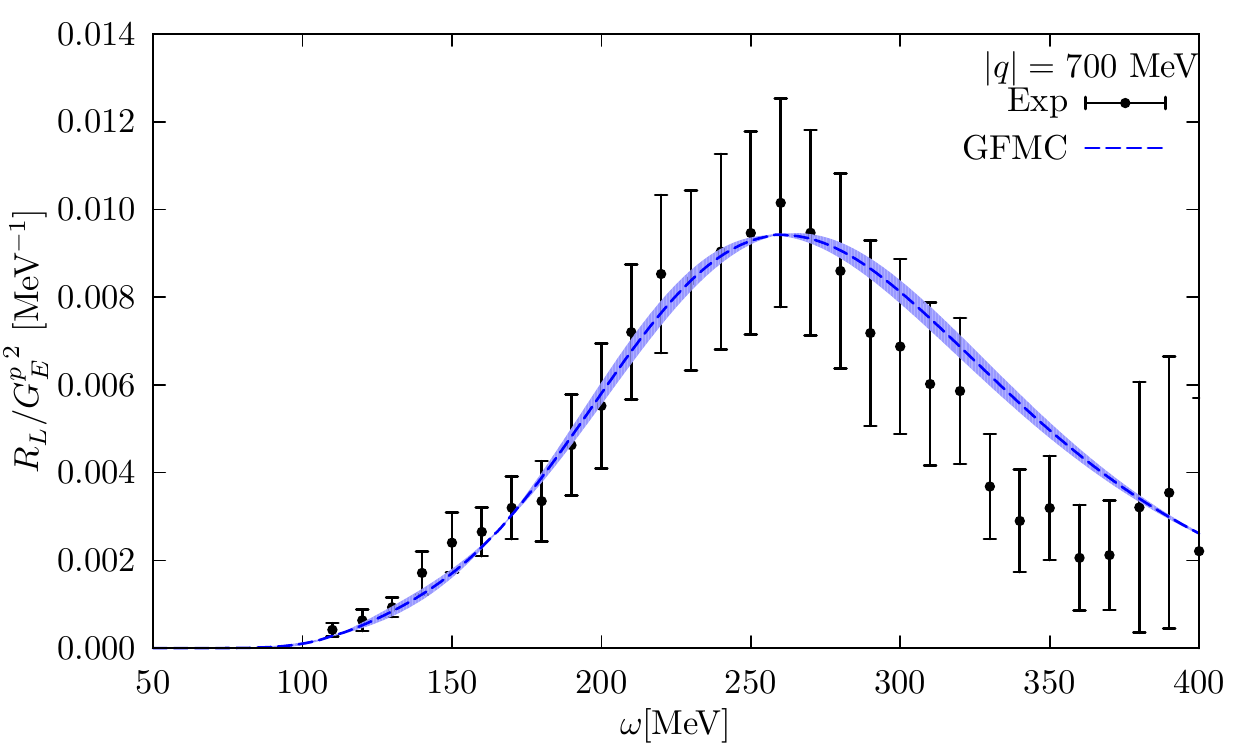}
\caption{ GFMC longitudinal electromagnetic response function of $^4$He at $|{\bf q}|$=700 MeV. 
Experimental data are from Ref.~\cite{Carlson:2002}.}
\label{fig:rl_700}
\end{figure}

\section{Inclusion of relativistic effects }\label{sec:rel:eff}
In Fig.~\ref{fig:rl_700} we compare the GFMC longitudinal response function of $^4$He divided by the 
proton electric form factor of Ref.~\cite{Hohler:1976ax} squared
with the corresponding experimental data for $|{\bf q}|=700$ MeV. We notice a slight shift of the  position 
of the quasielastic peak to higher $\omega$ and an overestimation of its width.
Here one has to take into account that although relativistic corrections up to order $q^2/m^2$, 
where $m$ is the mass of the nucleon, are included in the current operator for $R_L$, the quantum mechanical 
approach -- and hence the kinematics -- is nonrelativistic.
Strategies allowing to tackle relativistic corrections do exist in mean field approaches 
\cite{Amaro:2005dn}, however, an inclusion of relativistic effects in a fully interacting nuclear
many-body system is highly non-trivial. 
In order to cope with this problem, in the following, we will use the approach mentioned 
in the Introduction. 

In Refs.~\cite{Efros_LOT:2005,Efros_LOT:2010,Yuan:2010gh,Efros_LOT:2011,Yuan:2011rd}, 
it was proposed that one should perform the nonrelativistic 
calculation in a specific reference frame, where relativistic effects are as small as possible. 
For example, in electron nucleon scattering one prefers the Breit system, where the
initial nucleon is moving with ${\bf -q}/2$. A generalization to the quasielastic region
in electron nucleus scattering, which is dominated by a one-nucleon knock-out,
leads to 
the so-called active nucleon Breit (ANB) frame, where the target nucleus
moves with a momentum of -$A\,{\bf q}/2$. In this frame, any of the $A$ nucleons  composing the nucleus  
in the initial state has a momentum of about $-{\bf q}/2$, while the knocked-out nucleon carries a momentum 
$\simeq {\bf q}/2$ after the reaction.
In any other reference frame the involved momenta are higher. For example, in the laboratory (LAB) system
the knocked-out nucleon has a momentum of about ${\bf q}$, hence relativistic
effects can be minimized using the ANB system. 

Since experiments are carried out in the 
LAB system, it is necessary to transform the results from the ANB  (or any other frame where one performs the nonrelativistic
calculation) to the LAB frame. 
For reference frames moving  with respect to the LAB frame along the {\bf q} direction, as it is the case for the 
ANB frame, the responses transform as follows
\begin{align}\label{rllab}
R_L(|{\bf q}|,\omega)&= {\frac{{\bf q}^2}{({\bf q}^{\rm fr})^2}} 
{\frac{\sqrt{M_T^2+({\bf P}_i^{\rm fr})^2}}{M_T}} R_L(|{\bf q}^{\rm fr}|,\omega^{\rm fr}),\\
R_T(|{\bf q}|,\omega)&={\frac{\sqrt{M_T^2+({\bf P}_i^{\rm fr})^2}}{M_T}} R_T(|{\bf q}^{\rm fr}|,\omega^{\rm fr}) \,.\label{rtlab}
\end{align}
In the above equations $M_T$ is the mass of the target nucleus while $|{\bf q}^{\rm fr}|$ and $\omega^{\rm fr}$
are the momentum transfer and the energy transfer pertaining to the reference frame under consideration, namely 
\begin{align}
{\bf q}^{\rm fr}&={\bf P}_f^{\rm fr}-{\bf P}_i^{\rm fr}\nonumber\\
\omega^{\rm fr}&={ E}_f^{\rm fr}-{ E}_i^{\rm fr}
\label{rel_kin2}
\end{align}
where the total nonrelativistic energies ${ E}_{i/f}^{\rm fr}$ are given by
\begin{align}
{E}_i^{\rm fr}&=\frac{({\bf P}_i^{\rm fr})^2}{2 M_T}+\epsilon_0\nonumber\\
{E}_f^{\rm fr}&=\frac{({\bf P}_f^{\rm fr})^2}{2 M_T}+\epsilon_f\,,
\label{separ}
\end{align}
with
$ {\bf P}_{i/f}^{\rm fr}$ indicating the center-of-mass momenta in the specified reference frame. Since in a nonrelativistic calculation 
the intrinsic system does not depend on the center of mass momentum,
the intrinsic energies $\epsilon_f$ and $\epsilon_0$ are assumed to be frame independent.
\begin{figure}[]
\centering
\includegraphics[scale=0.675]{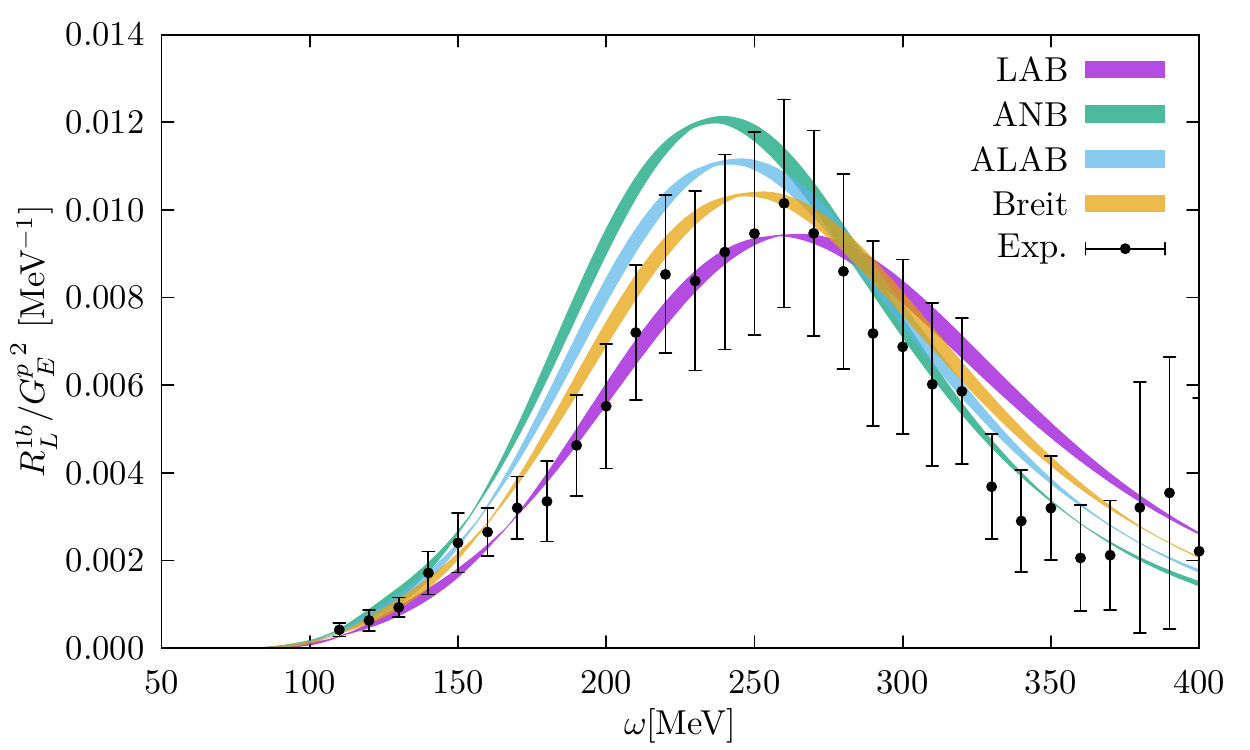}
\includegraphics[scale=0.675]{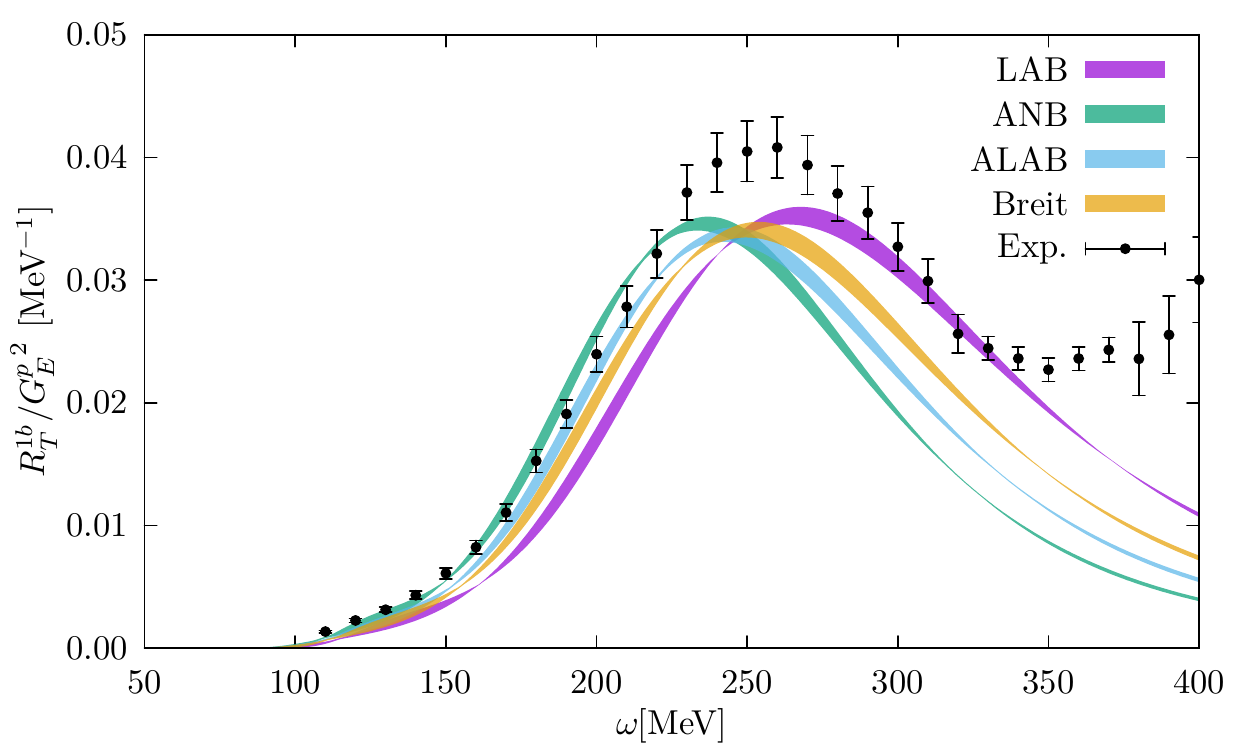}
\caption{ Frame dependence of the GFMC longitudinal (upper panel) and transverse (bottom panel) electromagnetic response functions of $^4$He at $|{\bf q}|$=700 MeV. }
\label{frame:dep}
\end{figure}
\begin{figure}[]
\centering
\includegraphics[scale=0.675]{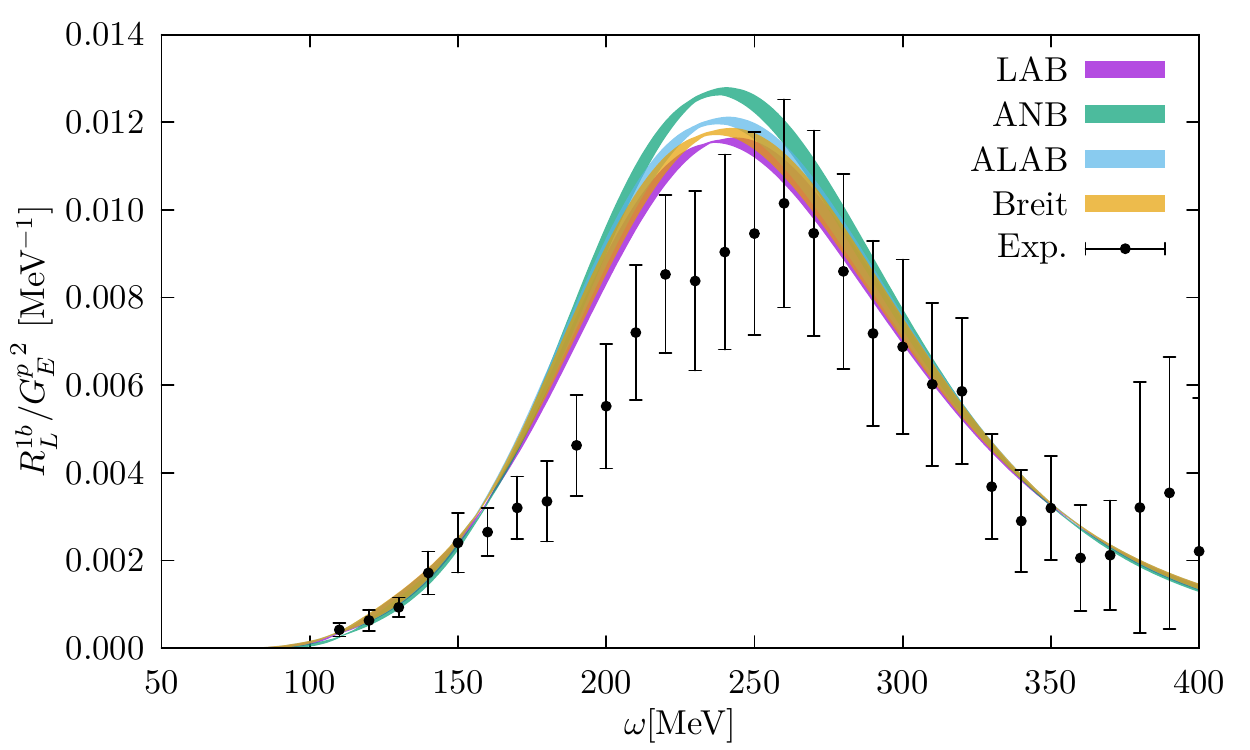}
\includegraphics[scale=0.675]{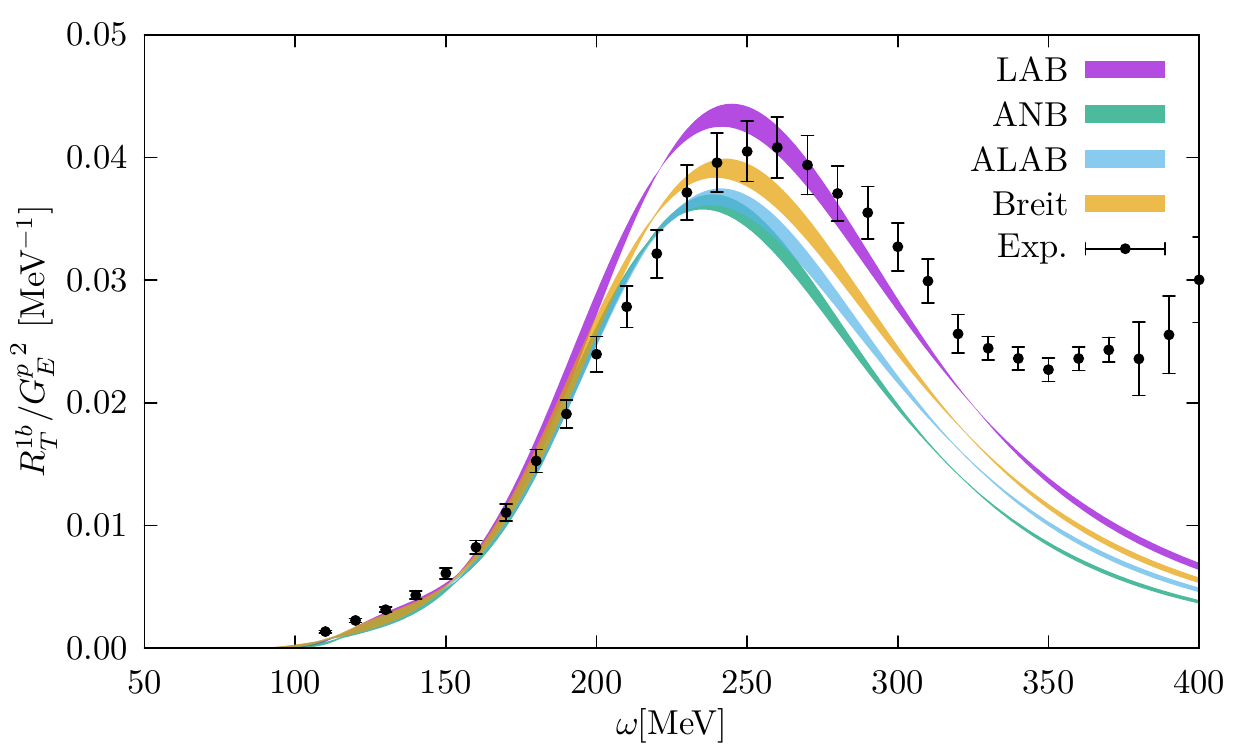}
\caption{ Same as Fig. \ref{frame:dep}, but considering two-body relativistic kinematics for the final state energy.  }
\label{frame:dep2}
\end{figure}

At first, we perform GFMC calculations for a number of momentum transfer of the intrinsic response functions, defined as
\begin{align}
&R^{\rm int}_\alpha( |{\bf q}^{\rm fr}|, \omega^{\rm int})=\sum_f \langle 0| j_{\alpha}^\dagger({\mathbf{q}^{\rm fr}},\omega^{\rm int}) |f\rangle   \nonumber\\
&\qquad \times \langle f | j_{\alpha}(\mathbf{q}^{\rm fr},\omega^{\rm int}) | 0\rangle \delta(\omega^{\rm int}-\epsilon_f + \epsilon_0)
\label{eq:res_int}
\end{align}
The direct calculation of the response functions in the LAB frame is simply achieved by taking ${\bf q}^{\rm fr}=\mathbf{q}$ 
and $\omega^{\rm int}=\omega-{\bf q}^{2}/(2 M_T)$.
On the other hand, in order to determine the responses in the LAB frame from Eqs.~(\ref{rllab})-(\ref{rtlab}),
$|{\bf q}|^{\rm fr}$ and $\omega^{\rm fr}$ are computed with the appropriate Lorentz
transformation from $|{\bf q}|$ and $\omega$. If $|{\bf q}^{\rm fr}|$ does not correspond to any of the tabulated momentum 
transfers, the response $R_{L/T}(|{\bf q}^{\rm fr}|,\omega^{\rm fr})$ is obtained interpolating the intrinsic response 
function using the procedure described in Sec.~\ref{sec:cross:sec} for  
$\omega^{\rm int}=\omega^{fr}-({\bf P}_f^{\rm fr})^2/(2 M_T)+ ({\bf P}_i^{\rm fr})^2/(2 M_T)$

\subsection*{The two-fragment model}\label{2frag:mod}

Relativistic effects in the kinematics can be included 
employing the two-fragment model of Ref.~\cite{Efros_LOT:2005}.
This relies on the assumption that the quasielastic reaction is dominated by the break-up
of the nucleus into two fragments, namely a knocked-out nucleon  
and a remaining $(A-1)$ system in its ground state. 
This assumption enables one to connect $\omega^{fr}$ to the intrinsic excitation energy $\epsilon_f$ 
used in the nonrelativistic calculation in a relativistically correct way.
It has to be noted that the two-fragment model is adopted only for determining the kinematic input of 
a calculation where the full nuclear dynamics of the system is taken into account.

At this point, we recall that within a nonrelativistic theory it is not possible to work 
simultaneously with the correct relativistic energy and momentum of a two-fragment system. 
As pointed out in \cite{Efros_LOT:2005}, a clue comes from the two-nucleon case. In fact, 
NN potential models are constructed describing the two-nucleon relative scattering momentum $p_{12}$ 
in a relativistically correct way, whereas the Schr\"odinger equation is solved for the ``fake'' nonrelativistic
 kinetic energy $E_{12}=p^2_{12}/2\mu_{12}$, where $\mu_{12}$ is the reduced mass of 
the two nucleons. (The same approach is also used in deuteron electrodisintegration, see, e.g. \cite{Ritz:1996za}).

Proceeding analogously to the NN potential case, the two-fragment kinematical model can be summarized by the
following points

a) The choice of the frame defines  ${\bf P}_i^{\rm fr}$, and accordingly also the initial relativistic hadron energy
\begin{equation}
 E_i^{\rm fr}=\sqrt{M_T^2+({\bf P}_i^{\rm fr})^2}
\end{equation}

b) The momenta of the knocked-out nucleon and the spectator system are set equal to
${\bf p}_N^{\rm fr}$ and ${\bf p}_X^{\rm fr}$, respectively. The corresponding relative and center-of-mass momenta are obtained as
\begin{eqnarray}
{\bf p}^{\rm fr}_{\rm f} &=& \mu({\frac{{\bf p}_N^{\rm fr}}{m}} - {\frac{{\bf p}_X^{\rm fr}}{M_X}}) \,,\\
{\bf P}_f^{\rm fr} &=& {\bf p}_N^{\rm fr} + {\bf p}_X^{\rm fr} \,,
\end{eqnarray}
where $M_X$ and $\mu$ are the mass of the spectator system and the
reduced mass, respectively; 

c) for reference frames moving with respect to the LAB frame along the ${\bf q}^{\rm fr}$ direction,
${\bf P}_f^{\rm fr}$
is directed along ${\bf q}^{\rm fr}$. In addition, for a quasielastic reaction 
one can safely assume that also ${\bf p}^{\rm fr}$ is directed along ${\bf q}^{\rm fr}$. Therefore 
${\bf p}^{\rm fr}_f$ and ${\bf P}^{\rm fr}_f$ have the same direction.
Under this assumption, ${ p}^{\rm fr}_f$  can be obtained from the relativistically correct final 
state energy of the hadron system
\begin{eqnarray}
\label{rel_kin}\nonumber
E_f^{fr} &=& \sqrt{m^2 + ({\bf p}^{\rm fr}_f + (\mu/M_{A-1}){\bf P}_f^{\rm fr})^2} \\
         &+&  \sqrt{M_{A-1}^2 + ({\bf p}^{\rm fr}_f - (\mu/m){\bf P}_f^{\rm fr})^2} \,;
\end{eqnarray}

 d) for each value of $\omega^{\rm fr}$ and $q^{\rm fr}$, one obtains $P_f^{\rm fr}$ 
and $E_f^{\rm fr}$ from Eq.(\ref{rel_kin2}). The relativistic relative momentum of the two fragments
is determined plugging Eq.~\eqref{rel_kin} into Eq.~\eqref{rel_kin2}. This then leads to the determination
of the intrinsic energy
\begin{equation}
\epsilon_f= \frac{(p_f^{\rm fr})^2}{2\mu}+\epsilon_0^{A-1}\,,
\end{equation}
where $(p_f^{\rm fr})^2/2\mu$ is the relativistically ``fake'' kinetic energy and $\epsilon_0^{A-1}$ the ground-state
energy of the spectator system. Finally, the response function of the two-body 
fragment model can be computed interpolating the intrinsic response of Eq.~(\ref{eq:res_int}) at
\begin{equation} 
\omega^{\rm int}=\frac{(p_f^{\rm fr})^2}{2\mu} -\epsilon_0+\epsilon_0^{A-1}\, .
\end{equation}

As further discussed in \cite{Efros_LOT:2005} one also has  to rescale the response functions
(see Eqs.~(9)-(11) therein). At this point one transforms the results to the LAB system as described 
above for the case without the two-fragment model.

Using the LIT method, the two-fragment model has been applied to the calculation of 
the $^3$He longitudinal~\cite{Efros_LOT:2005} and transverse response functions~\cite{Efros_LOT:2011,Yuan:2011rd}. 
Meson exchange and $\Delta$ isobar currents as well as relativistic corrections of order 
${\bf q}^2/m^2$ for the one-body charge and current operators were included. There it was shown 
that the large frame dependence of the results is almost eliminated by the use of the two-fragment relativistic kinematics,
 even considering momentum transfers up to 
$|{\bf q}|=700$  MeV. In particular a considerable shift of the quasielastic peak was found for all reference frames but the ANB one.
An excellent description of experimental $^3$He$(e,e')$ data 
is achieved when making the calculation in the ANB frame supplementing it with the two-fragment model.

In this work, we have calculated the GFMC electromagnetic responses of the four-body system in the same reference frames 
as in~\cite{Efros_LOT:2005,Yuan:2011rd,Efros_LOT:2011}. For $|{\bf q}|$=700 MeV we obtain the results 
shown in Fig.~\ref{frame:dep} for the longitudinal (upper panel) and transverse (lower panel) channel, respectively.
As in the three-body case, a rather strong frame dependence can be noticed, indicating that 
relativistic effects play a non negligible role at this value of the momentum transfer. 
The corresponding results obtained employing the two-fragment model are displayed in Fig.~\ref{frame:dep2}.
The position of the quasielastic peak of the electromagnetic responses no longer depends upon the reference frame and 
coincides with that of the ANB frame of Fig.~\ref{frame:dep}. 
Whilst in the longitudinal channel the different curves are almost coincident, the transverse responses still suffer a residual frame 
dependence, leading to different heights of the quasielastic peak. This has to be ascribed to the fact that, at variance
with Ref.~\cite{Efros_LOT:2011}, the sub-leading relativistic corrections in the transverse current operator are neglected
in the GFMC calculations. Our results are consistent with the findings of Ref.~\cite{Rocco:2016ejr}, where the role of relativistic effects in 
the kinematics and in the current operator is separately analyzed. In the LAB frame using relativistic currents brings about a 
reduction of the strength of the transverse response compared to the nonrelativistic ones. This effect, is expected to be smaller 
in the ANB frame, where the $\omega$-dependent correction in the current considered in Ref.~\cite{Efros_LOT:2011} vanishes 
at the quasielastic peak.  

There is a fairly good agreement between theory and experiment for the position of the quasielastic peak, in both the longitudinal
and the transverse channels. As for the peak heights, in the longitudinal case our calculations slightly overestimate the experimental
data, consistently with Ref.~\cite{Efros_LOT:2005} for the $^3$He case. In the transverse channel, for the afore-mentioned missing 
relativistic corrections in the current operator, only the ANB predictions can be meaningfully compared with experiments. Here, excess
strength from meson-exchange two-body currents is needed to bring GFMC results in agreement with experiments even in the 
quasielastic peak region. 

\begin{figure}[b]
\centering
\includegraphics[scale=0.675]{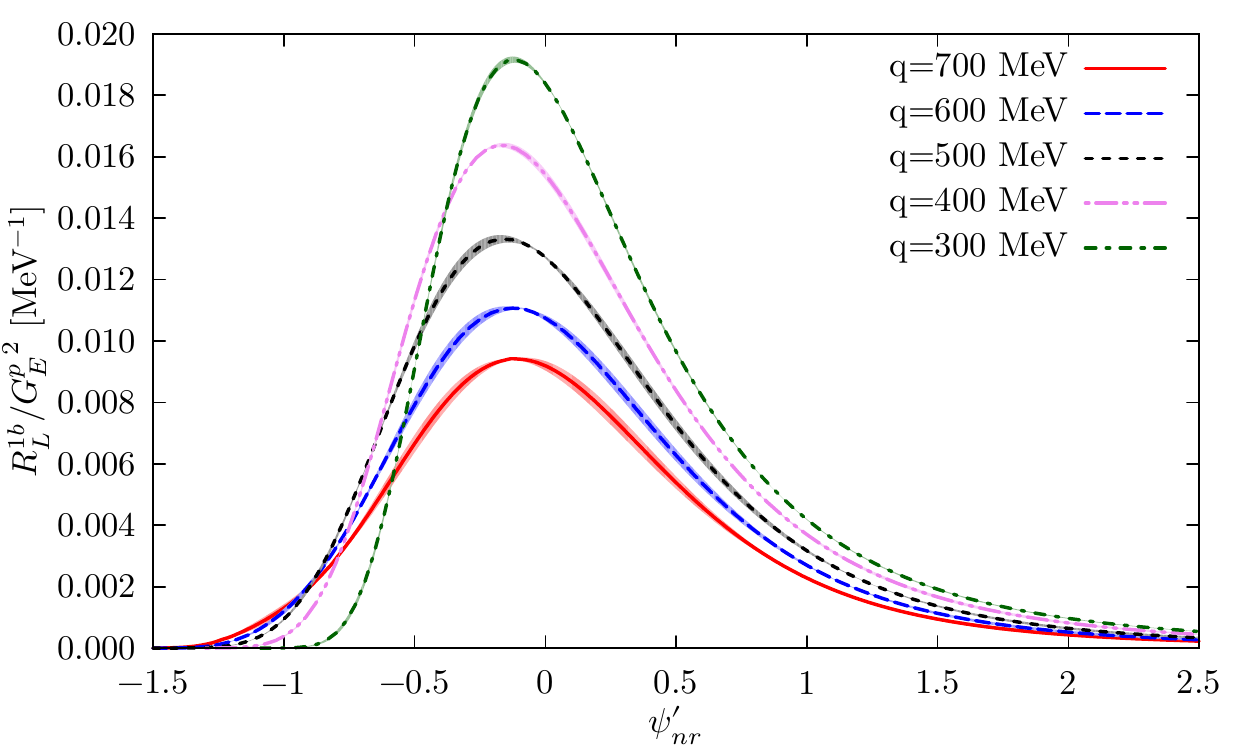}
\includegraphics[scale=0.675]{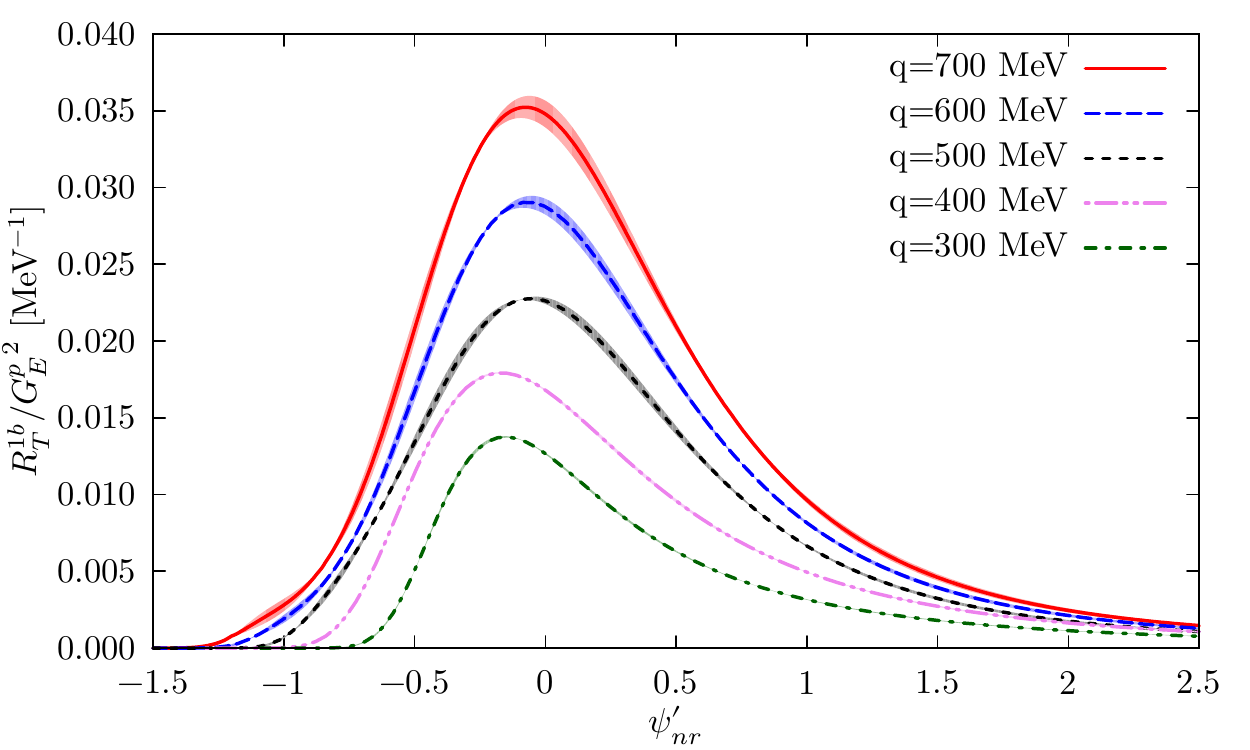}
\caption{ One-body longitudinal (upper panel) and transverse (bottom panel) electromagnetic response 
functions of $^4$He for $|{\bf q}_i|=300,400,500,600,$ and $700$ MeV as a function $\psi^\prime_{nr}$ given in Eq.\eqref{psi:nr} . }
\label{he4:psi:1b}
\end{figure}

\section{From response functions to cross sections}
\label{sec:cross:sec}
The calculation of the inclusive electron-nucleus scattering cross section of Eq.~\eqref{eq:x:sec}, 
requires the knowledge of $R_L$ and $R_T$ for several values of $\omega$ and $|{\bf q}|$. 
Hence, due to the sizable computational effort required to accurately invert the Euclidean response
for a given value of $|{\bf q}|$, the direct evaluation Eq.~\eqref{eq:x:sec} is not feasible within GFMC. 
To circumvent these difficulties, we developed a novel interpolation algorithm based on the scaling 
of the nuclear responses. The latter has been introduced and widely analyzed in the framework of 
the Global Relativistic Fermi gas (GRFG) model~\cite{Alberico:1988bv,Barbaro:1998gu}. Scaling of
the first kind occurs when the response functions divided by an appropriate factor, which accounts
for single-nucleon physics, no longer depend on ${\bf q}$ and $\omega$, but only upon a specific
function of them, which defines the scaling variable $\psi$.
Recently, the Authors of Ref. \cite{Rocco:2017hmh} carried out an analysis of the scaling features
of the GFMC electromagnetic response functions of $^4$He and $^{12}$C, retaining only one-body
current contributions. Their results show that scaling is fulfilled, provided that the nonrelativistic scaling
variable $\psi^{nr}$ is used. The latter is obtained from the nonrelativistic reduction of the energy-conserving
delta function of Eq.~\eqref{eq:RLT}, assuming that  the scattering process takes place on a single nucleon
and using the free energy spectrum for the initial and final states. 
In this work, we introduce a constant shift in the energy transfer in the definition of the scaling variable  
\begin{align}
\psi^\prime_{nr}= p_F\Big(\frac{\omega-E_{s}}{|{\bf q}|}-\frac{|{\bf q}|}{2m}\Big)\ .
\label{psi:nr}
\end{align}
In the above equation, $p_F$ is the Fermi momentum, and $E_s$ is empirically chosen to account for
binding effects in both the initial and final states. In the present analysis of the $^4$He nucleus, we use
$p_F$=180 MeV and $E_s=15$ MeV. However the results are quite insensitive to small variations of
these parameters. 

\begin{figure}[h]
\centering
\includegraphics[scale=0.675]{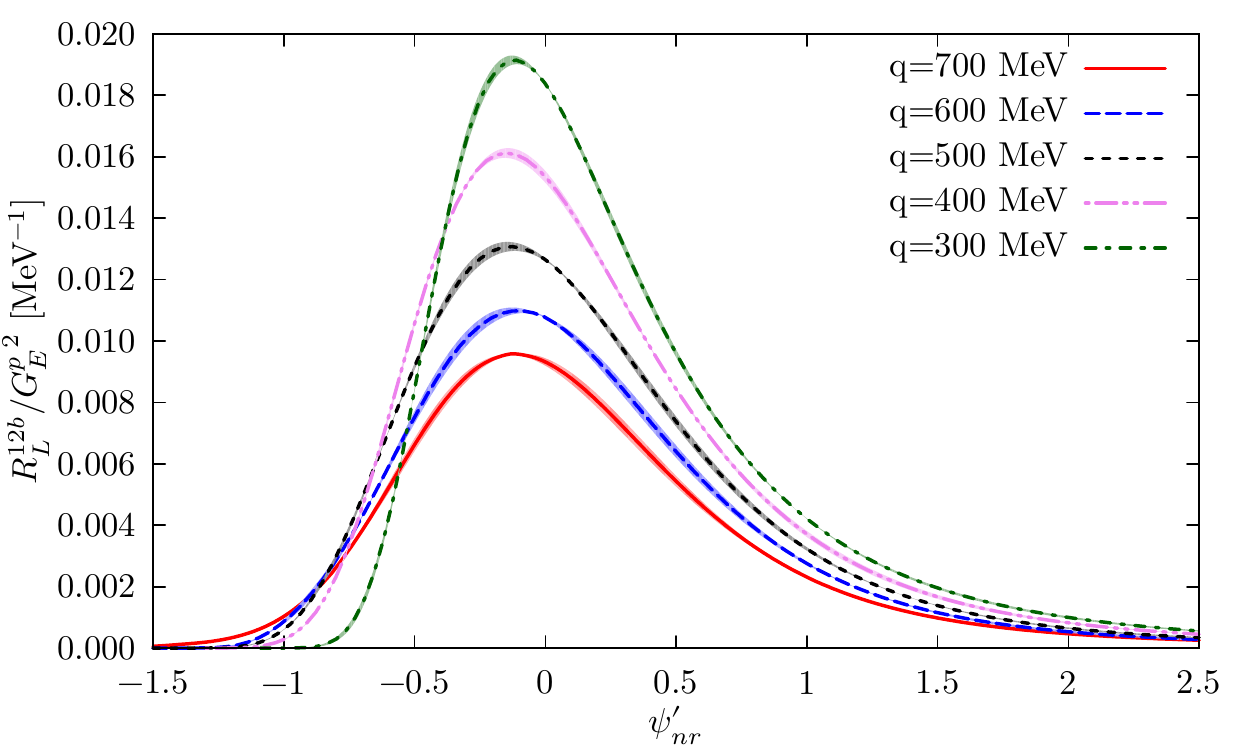}
\includegraphics[scale=0.675]{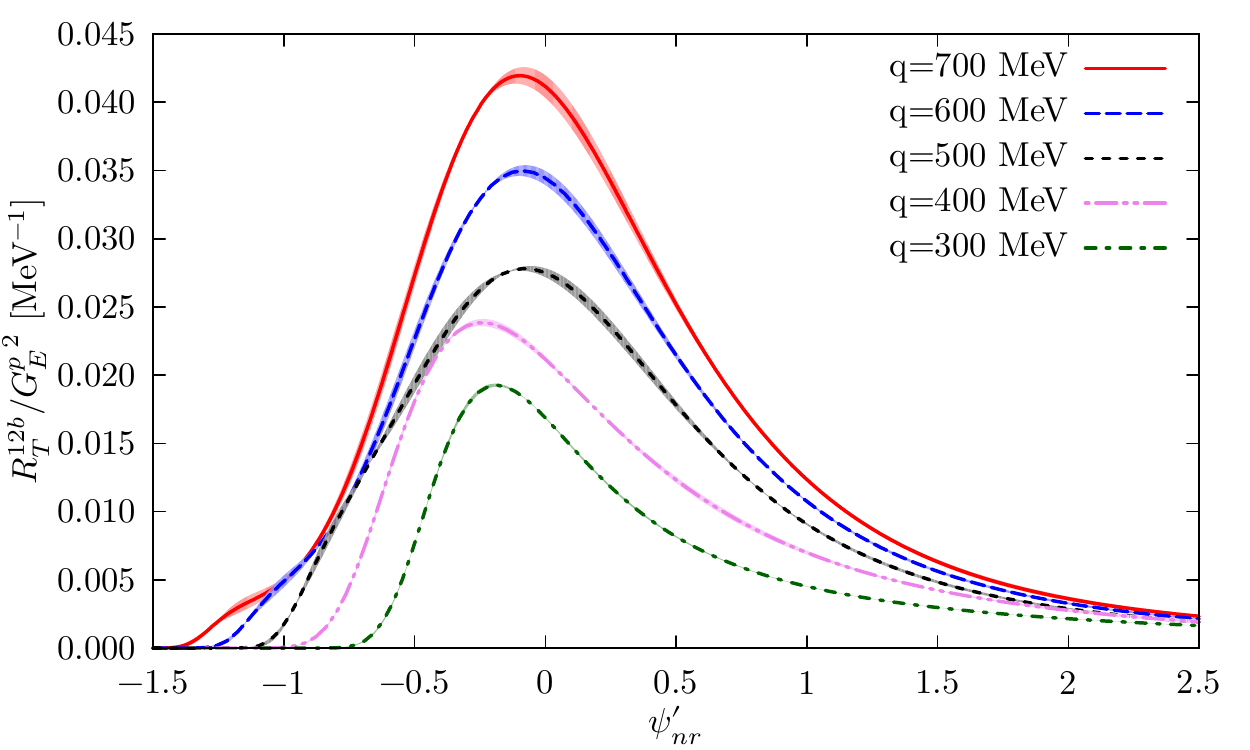}
\caption{ Same as in Fig.~\ref{he4:psi:1b} including one- and two-body terms in the electromagnetic current. }
\label{he4:psi:12b}
\end{figure}

Figure~\ref{he4:psi:1b} shows the longitudinal and transverse response functions of $^4$He divided by the 
proton electric form factor squared for $|{\bf q}_i|=300,400,500,600,$ and $700$ MeV as a function 
$\psi^\prime_{nr}$.
In both channels the curves corresponding to different values of the momentum transfer peak around $\psi^\prime_{nr}$=0
and the height of the quasielastic peaks is a monotonic function of $|{\bf q}|$. In the longitudinal case, shown in
the upper panel, the highest and the lowest peak correspond to $|{\bf q}|=300$ and 700 MeV, respectively. 
On the other hand, in the transverse channel, displayed in the bottom panel, the response functions are
smaller as $|{\bf q}|$ decreases. In Fig.~\ref{he4:psi:12b} both one- and two-body terms in the electromagnetic
current have been included. Meson-exchange current contributions only appreciably affect the transverse channel, leading 
to a sizable enhancement of the response functions. Nevertheless, the behavior of the curves in both the upper 
and lower panels is analogous to that of Fig.~\ref{he4:psi:1b}.

In order to evaluate Eq.~\eqref{eq:x:sec} we fix $E_e$ and $\theta_e$, the initial electron beam energy and 
scattering angle, respectively, and use $E_{e^\prime}=E_e-\omega$ for the energy of the outgoing electron. 
The four-momentum transfer is then written as
\begin{align}
Q^2=-q^2=4E_e(E_e-\omega)\sin^2\frac{\theta_e}{2}\, .
\end{align}
For a given value of $\omega$,  the response functions have to be evaluated at $|{\bf q}|=\sqrt{\omega^2+Q^2}$.
To this aim, we first compute $\psi^\prime_{nr}$ as in Eq.\eqref{psi:nr}. Then, the set of $R_{L,T}(\psi^\prime_{nr},q_i)$ 
is interpolated at $|{\bf q}|$. By looking at Figs.~\ref{he4:psi:1b} and \ref{he4:psi:12b}, it becomes evident why it is 
more convenient to interpolate the different response functions when the latter are given as a function of 
$\psi^\prime_{nr}$ and $|{\bf q}|$ rather than $\omega$ and $|{\bf q}|$. For a given value of $\psi^\prime_{nr}$ the 
curves corresponding to the different $|{\bf q}_i|$ are indeed almost perfectly aligned and monotonic functions of 
$|{\bf q}|$, largely improving the accuracy of the interpolation procedure. 

\begin{figure*}[h]
\includegraphics[scale=0.95]{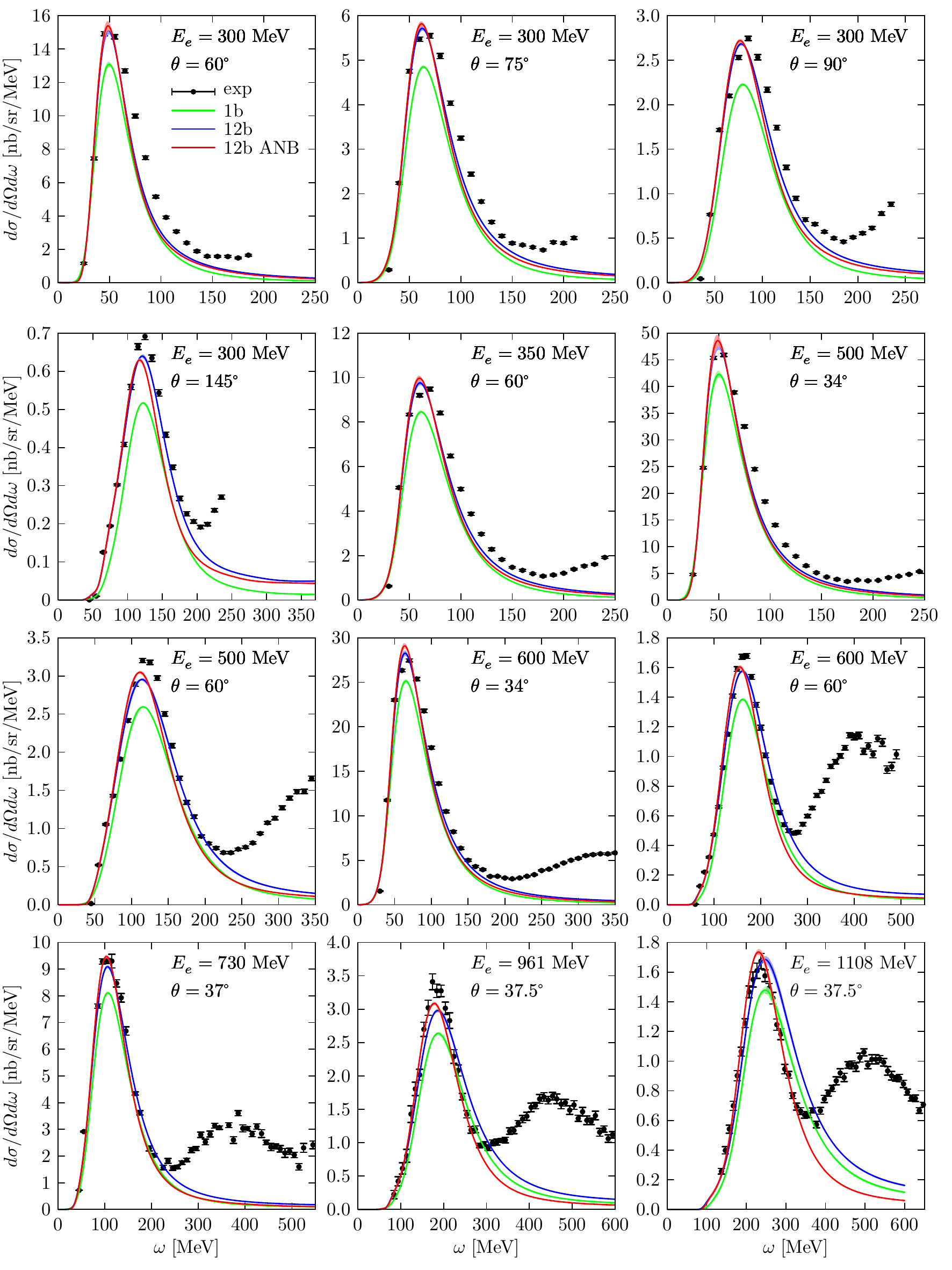}
\caption{Double-differential electron-$^4$He cross sections for different values of incident electron energy 
and scattering angle. The green and blue lines correspond to GFMC calculation were only one- body and
one- plus two-body contributions in the electromagnetic currents are accounted for. The red line indicates
one plus two-body current results obtained in the ANB frame, employing the two-body fragment model to 
account for relativistic kinematics. The experimental data are taken from Ref.~\cite{Benhar:2006er}. }
\label{cross:sec}
\end{figure*}

In Fig.~\ref{cross:sec} we compare with experimental data the electron-$^4$He inclusive double-differential
cross sections obtained from the GFMC responses for various kinematic setups, corresponding to different
values of $E_e$ and $\theta_e$. The green and blue curves correspond to retaining only one-body terms or 
both one- and two-body terms in the current operators. The red curves--which accounts for the contribution of 
one- plus two-body current operators--have been obtained computing the cross section in the ANB frame 
employing the two-fragment model to account for relativistic kinematics, and boosting back to the LAB frame. 

Our findings are consistent with those of Ref.~\cite{Lovato:2015}, as we observe that the two-body currents generate a large 
excess of strength over the whole $\omega$ spectrum largely improving the agreement with experimental data. 
The difference between the red and blue curves is clearly visible for $E_e= 961,\ 1080$ MeV and $\theta_e=37.5^\circ$,
where $Q^2\gtrsim 3$ GeV$^2$ at the quasielastic peak. In these two kinematic setups, the inclusion of relativistic
corrections lead to a shift in the position of the quasielastic peak and a reduction of its width. The latter effect is needed
to not overestimate the experimental data once the resonance production mechanism is accounted for.

\section{Conclusions}
\label{sec:concl}
The electromagnetic longitudinal responses of $^4$He obtained with the GFMC have been successfully benchmarked with some 
LIT results from the literature~\cite{Bacca:2009,Bacca:2009prc}
 For 
$|{\bf q}|$=300 MeV we have found a very good agreement between the two theoretical ab initio approaches.  We have checked that 
the small discrepancies,
which become more pronounced at $|{\bf q}|$=500 MeV, are not due to problems pertaining to the inversion procedure. 
They are likely to be ascribed to various smaller differences in the calculations listed in 
Section~\ref{IIB}. 

We have gauged the relativistic effects in the GFMC electromagnetic response functions at relatively high value of the 
momentum transfer, $|\mathbf{q}|=700$ MeV. To this aim, we have computed the response functions in different reference
frames, boosting the results back to the LAB frame. We observe sizable differences in the position and the strength of
the quasielastic peak. The two-fragment model of Ref.~\cite{Efros_LOT:2005}, suitable for realistic models of nuclear dynamics,  
has been employed to account for relativistic kinematics. This method has proven to provide fully satisfactory results in the
longitudinal channel. As for the transverse channel, residual frame dependence in the strength of the quasielastic peak are
likely to be due to the missing higher-order relativistic corrections in the transition operator. This is consistent with the findings
of Refs.~\cite{Efros_LOT:2011,Rocco:2016ejr} and their inclusion will be the subject of future work. 

A novel algorithm to reliably and efficiently interpolate the GFMC response functions for arbitrary values of $|\mathbf{q}|$
and $\omega$ has been devised. This algorithm relies on the first-kind scaling features of the GFMC responses, which 
has been analyzed in Ref.~\cite{Rocco:2017hmh}. It has to be noted that scaling violations do not prevent its 
application.  On the other hand, if scaling were exactly fulfilled, 
the algorithm would only require the GFMC calculation of the
response functions for a single value of $|\mathbf{q}|$. 

We have employed the interpolation algorithm to perform the first 
\textit{ab-initio} calculation of 
the double differential cross section
of the inclusive electron-$^4$He scattering. The extensive comparison with experimental data demonstrates that two-body 
currents generate an excess of strength that is necessary to correct the cross section, even in the quasielastic peak region. 
Relativistic corrections, only appreciable for larger values of the lepton energy and scattering angles, lead to a shift in the position 
of the quasielastic peak and a reduction of its width. Our findings indicate that relativistic effects are primarily kinematical in nature,
and can easily be accounted for in the GFMC or any nuclear \textit{ab-initio} approach, provided that the many-body calculations are 
carried out in a proper reference frame.
Therefore the fact that neutrino fluxes in current and planned experiments cover a
broad energy range extending to several GeVs does not invalidate per se the results obtained within the nonrelativistic approach.

\section{Acknowledgements} 
Many illuminating discussions with S. Bacca, J. Carlson, J. Nieves and R. Schiavilla are gratefully acknowledged. 
This research has been supported by the U.S. Department of Energy, Office of Science, Office of Nuclear Physics, under contract DE-AC02-06CH11357, and by the Centro 
Nazionale delle Ricerche (CNR) and the Royal Society under the CNR-Royal Society International Fellowship scheme NF161046. Under 
an award of computer time provided by the INCITE program, this research used resources of the Argonne Leadership Computing Facility
at Argonne National Laboratory, which is supported by the Office of Science of the U.S. Department of Energy under contract DE-AC02-06CH11357.

\bibliography{biblio}

\begin{thebibliography}{47}%
\makeatletter
\providecommand \@ifxundefined [1]{%
 \@ifx{#1\undefined}
}%
\providecommand \@ifnum [1]{%
 \ifnum #1\expandafter \@firstoftwo
 \else \expandafter \@secondoftwo
 \fi
}%
\providecommand \@ifx [1]{%
 \ifx #1\expandafter \@firstoftwo
 \else \expandafter \@secondoftwo
 \fi
}%
\providecommand \natexlab [1]{#1}%
\providecommand \enquote  [1]{``#1''}%
\providecommand \bibnamefont  [1]{#1}%
\providecommand \bibfnamefont [1]{#1}%
\providecommand \citenamefont [1]{#1}%
\providecommand \href@noop [0]{\@secondoftwo}%
\providecommand \href [0]{\begingroup \@sanitize@url \@href}%
\providecommand \@href[1]{\@@startlink{#1}\@@href}%
\providecommand \@@href[1]{\endgroup#1\@@endlink}%
\providecommand \@sanitize@url [0]{\catcode `\\12\catcode `\$12\catcode
  `\&12\catcode `\#12\catcode `\^12\catcode `\_12\catcode `\%12\relax}%
\providecommand \@@startlink[1]{}%
\providecommand \@@endlink[0]{}%
\providecommand \url  [0]{\begingroup\@sanitize@url \@url }%
\providecommand \@url [1]{\endgroup\@href {#1}{\urlprefix }}%
\providecommand \urlprefix  [0]{URL }%
\providecommand \Eprint [0]{\href }%
\providecommand \doibase [0]{http://dx.doi.org/}%
\providecommand \selectlanguage [0]{\@gobble}%
\providecommand \bibinfo  [0]{\@secondoftwo}%
\providecommand \bibfield  [0]{\@secondoftwo}%
\providecommand \translation [1]{[#1]}%
\providecommand \BibitemOpen [0]{}%
\providecommand \bibitemStop [0]{}%
\providecommand \bibitemNoStop [0]{.\EOS\space}%
\providecommand \EOS [0]{\spacefactor3000\relax}%
\providecommand \BibitemShut  [1]{\csname bibitem#1\endcsname}%
\let\auto@bib@innerbib\@empty
\bibitem [{The MINER$\nu$A Experiment()}]{minerva_web}%
  \BibitemOpen
  The MINER$\nu$A Experiment,\ \href@noop {} {}\bibinfo {howpublished}
  {\url{http://minerva.fnal.gov}}\BibitemShut {NoStop}%
\bibitem [{The MicroBooNE Experiment()}]{mb_web}%
  \BibitemOpen
  The MicroBooNE Experiment,\ \href@noop {} {}\bibinfo {howpublished}
  {\url{http://www-microboone.fnal.gov}}\BibitemShut {NoStop}%
\bibitem [{The {NO}v{A} {E}xperiment()}]{nova_web}%
  \BibitemOpen
  The {NO}v{A} {E}xperiment,\ \href@noop {} {}\bibinfo {howpublished}
  {\url{http://www-nova.fnal.gov}}\BibitemShut {NoStop}%
\bibitem [{The {T}2{K} {E}xperiment()}]{t2k_web}%
  \BibitemOpen
  The {T}2{K} {E}xperiment,\ \href@noop {} {}\bibinfo {howpublished}
  {\url{http://t2k-experiment.org}}\BibitemShut {NoStop}%
\bibitem [{The Deep Underground Neutrino Experiment()}]{dune_web}%
  \BibitemOpen
  The Deep Underground Neutrino Experiment,\ \href@noop {} {}\bibinfo
  {howpublished} {\url{http://www.dunescience.org}}\BibitemShut {NoStop}%
\bibitem [{Hyper-Kamiokande()}]{hk_web}%
  \BibitemOpen
  Hyper-Kamiokande,\ \href@noop {} {}\bibinfo {howpublished}
  {\url{http://www.hyperk.org}}\BibitemShut {NoStop}%
\bibitem [{\citenamefont {Katori}\ and\ \citenamefont
  {Martini}(2018)}]{Katori:2016yel}%
  \BibitemOpen
  \bibfield  {author} {\bibinfo {author} {\bibfnamefont {T.}~\bibnamefont
  {Katori}}\ and\ \bibinfo {author} {\bibfnamefont {M.}~\bibnamefont
  {Martini}},\ }\href {\doibase 10.1088/1361-6471/aa8bf7} {\bibfield  {journal}
  {\bibinfo  {journal} {J. Phys.}\ }\textbf {\bibinfo {volume} {G45}},\
  \bibinfo {pages} {013001} (\bibinfo {year} {2018})},\ \Eprint
  {http://arxiv.org/abs/1611.07770} {arXiv:1611.07770 [hep-ph]} \BibitemShut
  {NoStop}%
\bibitem [{\citenamefont {Benhar}\ \emph {et~al.}(2017)\citenamefont {Benhar},
  \citenamefont {Huber}, \citenamefont {Mariani},\ and\ \citenamefont
  {Meloni}}]{Benhar:2015wva}%
  \BibitemOpen
  \bibfield  {author} {\bibinfo {author} {\bibfnamefont {O.}~\bibnamefont
  {Benhar}}, \bibinfo {author} {\bibfnamefont {P.}~\bibnamefont {Huber}},
  \bibinfo {author} {\bibfnamefont {C.}~\bibnamefont {Mariani}}, \ and\
  \bibinfo {author} {\bibfnamefont {D.}~\bibnamefont {Meloni}},\ }\href
  {\doibase 10.1016/j.physrep.2017.07.004} {\bibfield  {journal} {\bibinfo
  {journal} {Phys. Rept.}\ }\textbf {\bibinfo {volume} {700}},\ \bibinfo
  {pages} {1} (\bibinfo {year} {2017})},\ \Eprint
  {http://arxiv.org/abs/1501.06448} {arXiv:1501.06448 [nucl-th]} \BibitemShut
  {NoStop}%
\bibitem [{\citenamefont {Leidemann}\ and\ \citenamefont
  {Orlandini}(2013)}]{Lei13}%
  \BibitemOpen
  \bibfield  {author} {\bibinfo {author} {\bibfnamefont {W.}~\bibnamefont
  {Leidemann}}\ and\ \bibinfo {author} {\bibfnamefont {G.}~\bibnamefont
  {Orlandini}},\ }\href@noop {} {\bibfield  {journal} {\bibinfo  {journal}
  {Prog. Part. Nucl. Phys.}\ }\textbf {\bibinfo {volume} {68}},\ \bibinfo
  {pages} {158} (\bibinfo {year} {2013})}\BibitemShut {NoStop}%
\bibitem [{\citenamefont {Barrett}\ \emph {et~al.}(2013)\citenamefont
  {Barrett}, \citenamefont {Navratil},\ and\ \citenamefont
  {Vary}}]{Barrett:2013nh}%
  \BibitemOpen
  \bibfield  {author} {\bibinfo {author} {\bibfnamefont {B.~R.}\ \bibnamefont
  {Barrett}}, \bibinfo {author} {\bibfnamefont {P.}~\bibnamefont {Navratil}}, \
  and\ \bibinfo {author} {\bibfnamefont {J.~P.}\ \bibnamefont {Vary}},\ }\href
  {\doibase 10.1016/j.ppnp.2012.10.003} {\bibfield  {journal} {\bibinfo
  {journal} {Prog. Part. Nucl. Phys.}\ }\textbf {\bibinfo {volume} {69}},\
  \bibinfo {pages} {131} (\bibinfo {year} {2013})}\BibitemShut {NoStop}%
\bibitem [{\citenamefont {Hagen}\ \emph {et~al.}(2014)\citenamefont {Hagen},
  \citenamefont {Papenbrock}, \citenamefont {Hjorth-Jensen},\ and\
  \citenamefont {Dean}}]{Hagen:2013nca}%
  \BibitemOpen
  \bibfield  {author} {\bibinfo {author} {\bibfnamefont {G.}~\bibnamefont
  {Hagen}}, \bibinfo {author} {\bibfnamefont {T.}~\bibnamefont {Papenbrock}},
  \bibinfo {author} {\bibfnamefont {M.}~\bibnamefont {Hjorth-Jensen}}, \ and\
  \bibinfo {author} {\bibfnamefont {D.~J.}\ \bibnamefont {Dean}},\ }\href
  {\doibase 10.1088/0034-4885/77/9/096302} {\bibfield  {journal} {\bibinfo
  {journal} {Rept. Prog. Phys.}\ }\textbf {\bibinfo {volume} {77}},\ \bibinfo
  {pages} {096302} (\bibinfo {year} {2014})},\ \Eprint
  {http://arxiv.org/abs/1312.7872} {arXiv:1312.7872 [nucl-th]} \BibitemShut
  {NoStop}%
\bibitem [{\citenamefont {Hergert}\ \emph {et~al.}(2016)\citenamefont
  {Hergert}, \citenamefont {Bogner}, \citenamefont {Morris}, \citenamefont
  {Schwenk},\ and\ \citenamefont {Tsukiyama}}]{Hergert:2015awm}%
  \BibitemOpen
  \bibfield  {author} {\bibinfo {author} {\bibfnamefont {H.}~\bibnamefont
  {Hergert}}, \bibinfo {author} {\bibfnamefont {S.~K.}\ \bibnamefont {Bogner}},
  \bibinfo {author} {\bibfnamefont {T.~D.}\ \bibnamefont {Morris}}, \bibinfo
  {author} {\bibfnamefont {A.}~\bibnamefont {Schwenk}}, \ and\ \bibinfo
  {author} {\bibfnamefont {K.}~\bibnamefont {Tsukiyama}},\ }\href {\doibase
  10.1016/j.physrep.2015.12.007} {\bibfield  {journal} {\bibinfo  {journal}
  {Phys. Rept.}\ }\textbf {\bibinfo {volume} {621}},\ \bibinfo {pages} {165}
  (\bibinfo {year} {2016})},\ \Eprint {http://arxiv.org/abs/1512.06956}
  {arXiv:1512.06956 [nucl-th]} \BibitemShut {NoStop}%
\bibitem [{\citenamefont {Carbone}\ \emph {et~al.}(2013)\citenamefont
  {Carbone}, \citenamefont {Cipollone}, \citenamefont {Barbieri}, \citenamefont
  {Rios},\ and\ \citenamefont {Polls}}]{Carbone:2013eqa}%
  \BibitemOpen
  \bibfield  {author} {\bibinfo {author} {\bibfnamefont {A.}~\bibnamefont
  {Carbone}}, \bibinfo {author} {\bibfnamefont {A.}~\bibnamefont {Cipollone}},
  \bibinfo {author} {\bibfnamefont {C.}~\bibnamefont {Barbieri}}, \bibinfo
  {author} {\bibfnamefont {A.}~\bibnamefont {Rios}}, \ and\ \bibinfo {author}
  {\bibfnamefont {A.}~\bibnamefont {Polls}},\ }\href {\doibase
  10.1103/PhysRevC.88.054326} {\bibfield  {journal} {\bibinfo  {journal} {Phys.
  Rev.}\ }\textbf {\bibinfo {volume} {C88}},\ \bibinfo {pages} {054326}
  (\bibinfo {year} {2013})},\ \Eprint {http://arxiv.org/abs/1310.3688}
  {arXiv:1310.3688 [nucl-th]} \BibitemShut {NoStop}%
\bibitem [{\citenamefont {Epelbaum}\ \emph {et~al.}(2011)\citenamefont
  {Epelbaum}, \citenamefont {Krebs}, \citenamefont {Lee},\ and\ \citenamefont
  {Meissner}}]{Epelbaum:2011md}%
  \BibitemOpen
  \bibfield  {author} {\bibinfo {author} {\bibfnamefont {E.}~\bibnamefont
  {Epelbaum}}, \bibinfo {author} {\bibfnamefont {H.}~\bibnamefont {Krebs}},
  \bibinfo {author} {\bibfnamefont {D.}~\bibnamefont {Lee}}, \ and\ \bibinfo
  {author} {\bibfnamefont {U.-G.}\ \bibnamefont {Meissner}},\ }\href {\doibase
  10.1103/PhysRevLett.106.192501} {\bibfield  {journal} {\bibinfo  {journal}
  {Phys. Rev. Lett.}\ }\textbf {\bibinfo {volume} {106}},\ \bibinfo {pages}
  {192501} (\bibinfo {year} {2011})},\ \Eprint {http://arxiv.org/abs/1101.2547}
  {arXiv:1101.2547 [nucl-th]} \BibitemShut {NoStop}%
\bibitem [{\citenamefont {Carlson}\ \emph
  {et~al.}(2015{\natexlab{a}})\citenamefont {Carlson}, \citenamefont
  {Gandolfi}, \citenamefont {Pederiva}, \citenamefont {Pieper}, \citenamefont
  {Schiavilla}, \citenamefont {Schmidt},\ and\ \citenamefont
  {Wiringa}}]{Carlson:2014vla}%
  \BibitemOpen
  \bibfield  {author} {\bibinfo {author} {\bibfnamefont {J.}~\bibnamefont
  {Carlson}}, \bibinfo {author} {\bibfnamefont {S.}~\bibnamefont {Gandolfi}},
  \bibinfo {author} {\bibfnamefont {F.}~\bibnamefont {Pederiva}}, \bibinfo
  {author} {\bibfnamefont {S.~C.}\ \bibnamefont {Pieper}}, \bibinfo {author}
  {\bibfnamefont {R.}~\bibnamefont {Schiavilla}}, \bibinfo {author}
  {\bibfnamefont {K.~E.}\ \bibnamefont {Schmidt}}, \ and\ \bibinfo {author}
  {\bibfnamefont {R.~B.}\ \bibnamefont {Wiringa}},\ }\href {\doibase
  10.1103/RevModPhys.87.1067} {\bibfield  {journal} {\bibinfo  {journal} {Rev.
  Mod. Phys.}\ }\textbf {\bibinfo {volume} {87}},\ \bibinfo {pages} {1067}
  (\bibinfo {year} {2015}{\natexlab{a}})},\ \Eprint
  {http://arxiv.org/abs/1412.3081} {arXiv:1412.3081 [nucl-th]} \BibitemShut
  {NoStop}%
\bibitem [{\citenamefont {Hammer}\ \emph {et~al.}(2013)\citenamefont {Hammer},
  \citenamefont {Nogga},\ and\ \citenamefont {Schwenk}}]{Hammer:2012id}%
  \BibitemOpen
  \bibfield  {author} {\bibinfo {author} {\bibfnamefont {H.-W.}\ \bibnamefont
  {Hammer}}, \bibinfo {author} {\bibfnamefont {A.}~\bibnamefont {Nogga}}, \
  and\ \bibinfo {author} {\bibfnamefont {A.}~\bibnamefont {Schwenk}},\ }\href
  {\doibase 10.1103/RevModPhys.85.197} {\bibfield  {journal} {\bibinfo
  {journal} {Rev. Mod. Phys.}\ }\textbf {\bibinfo {volume} {85}},\ \bibinfo
  {pages} {197} (\bibinfo {year} {2013})},\ \Eprint
  {http://arxiv.org/abs/1210.4273} {arXiv:1210.4273 [nucl-th]} \BibitemShut
  {NoStop}%
\bibitem [{\citenamefont {Carlson}(1987)}]{Carlson:1987zz}%
  \BibitemOpen
  \bibfield  {author} {\bibinfo {author} {\bibfnamefont {J.}~\bibnamefont
  {Carlson}},\ }\href {\doibase 10.1103/PhysRevC.36.2026} {\bibfield  {journal}
  {\bibinfo  {journal} {Phys. Rev.}\ }\textbf {\bibinfo {volume} {C36}},\
  \bibinfo {pages} {2026} (\bibinfo {year} {1987})}\BibitemShut {NoStop}%
\bibitem [{\citenamefont {Lovato}\ \emph {et~al.}(2015)\citenamefont {Lovato},
  \citenamefont {Gandolfi}, \citenamefont {Carlson}, \citenamefont {Pieper},\
  and\ \citenamefont {Schiavilla}}]{Lovato:2015}%
  \BibitemOpen
  \bibfield  {author} {\bibinfo {author} {\bibfnamefont {A.}~\bibnamefont
  {Lovato}}, \bibinfo {author} {\bibfnamefont {S.}~\bibnamefont {Gandolfi}},
  \bibinfo {author} {\bibfnamefont {J.}~\bibnamefont {Carlson}}, \bibinfo
  {author} {\bibfnamefont {S.~C.}\ \bibnamefont {Pieper}}, \ and\ \bibinfo
  {author} {\bibfnamefont {R.}~\bibnamefont {Schiavilla}},\ }\href {\doibase
  10.1103/PhysRevC.91.062501} {\bibfield  {journal} {\bibinfo  {journal} {Phys.
  Rev. C}\ }\textbf {\bibinfo {volume} {91}},\ \bibinfo {pages} {062501}
  (\bibinfo {year} {2015})}\BibitemShut {NoStop}%
\bibitem [{\citenamefont {Lovato}\ \emph {et~al.}(2016)\citenamefont {Lovato},
  \citenamefont {Gandolfi}, \citenamefont {Carlson}, \citenamefont {Pieper},\
  and\ \citenamefont {Schiavilla}}]{Lovato:2016gkq}%
  \BibitemOpen
  \bibfield  {author} {\bibinfo {author} {\bibfnamefont {A.}~\bibnamefont
  {Lovato}}, \bibinfo {author} {\bibfnamefont {S.}~\bibnamefont {Gandolfi}},
  \bibinfo {author} {\bibfnamefont {J.}~\bibnamefont {Carlson}}, \bibinfo
  {author} {\bibfnamefont {S.~C.}\ \bibnamefont {Pieper}}, \ and\ \bibinfo
  {author} {\bibfnamefont {R.}~\bibnamefont {Schiavilla}},\ }\href {\doibase
  10.1103/PhysRevLett.117.082501} {\bibfield  {journal} {\bibinfo  {journal}
  {Phys. Rev. Lett.}\ }\textbf {\bibinfo {volume} {117}},\ \bibinfo {pages}
  {082501} (\bibinfo {year} {2016})},\ \Eprint
  {http://arxiv.org/abs/1605.00248} {arXiv:1605.00248 [nucl-th]} \BibitemShut
  {NoStop}%
\bibitem [{\citenamefont {Amaro}\ \emph {et~al.}(2005)\citenamefont {Amaro},
  \citenamefont {Barbaro}, \citenamefont {Caballero}, \citenamefont
  {Donnelly},\ and\ \citenamefont {Maieron}}]{Amaro:2005dn}%
  \BibitemOpen
  \bibfield  {author} {\bibinfo {author} {\bibfnamefont {J.~E.}\ \bibnamefont
  {Amaro}}, \bibinfo {author} {\bibfnamefont {M.~B.}\ \bibnamefont {Barbaro}},
  \bibinfo {author} {\bibfnamefont {J.~A.}\ \bibnamefont {Caballero}}, \bibinfo
  {author} {\bibfnamefont {T.~W.}\ \bibnamefont {Donnelly}}, \ and\ \bibinfo
  {author} {\bibfnamefont {C.}~\bibnamefont {Maieron}},\ }\href {\doibase
  10.1103/PhysRevC.71.065501} {\bibfield  {journal} {\bibinfo  {journal} {Phys.
  Rev.}\ }\textbf {\bibinfo {volume} {C71}},\ \bibinfo {pages} {065501}
  (\bibinfo {year} {2005})},\ \Eprint {http://arxiv.org/abs/nucl-th/0503062}
  {arXiv:nucl-th/0503062 [nucl-th]} \BibitemShut {NoStop}%
\bibitem [{\citenamefont {Efros}\ \emph {et~al.}(2005)\citenamefont {Efros},
  \citenamefont {Leidemann}, \citenamefont {Orlandini},\ and\ \citenamefont
  {Tomusiak}}]{Efros_LOT:2005}%
  \BibitemOpen
  \bibfield  {author} {\bibinfo {author} {\bibfnamefont {V.~D.}\ \bibnamefont
  {Efros}}, \bibinfo {author} {\bibfnamefont {W.}~\bibnamefont {Leidemann}},
  \bibinfo {author} {\bibfnamefont {G.}~\bibnamefont {Orlandini}}, \ and\
  \bibinfo {author} {\bibfnamefont {E.~L.}\ \bibnamefont {Tomusiak}},\ }\href
  {\doibase 10.1103/PhysRevC.72.011002} {\bibfield  {journal} {\bibinfo
  {journal} {Phys. Rev. C}\ }\textbf {\bibinfo {volume} {72}},\ \bibinfo
  {pages} {011002} (\bibinfo {year} {2005})}\BibitemShut {NoStop}%
\bibitem [{\citenamefont {Efros}\ \emph {et~al.}(2010)\citenamefont {Efros},
  \citenamefont {Leidemann}, \citenamefont {Orlandini},\ and\ \citenamefont
  {Tomusiak}}]{Efros_LOT:2010}%
  \BibitemOpen
  \bibfield  {author} {\bibinfo {author} {\bibfnamefont {V.~D.}\ \bibnamefont
  {Efros}}, \bibinfo {author} {\bibfnamefont {W.}~\bibnamefont {Leidemann}},
  \bibinfo {author} {\bibfnamefont {G.}~\bibnamefont {Orlandini}}, \ and\
  \bibinfo {author} {\bibfnamefont {E.~L.}\ \bibnamefont {Tomusiak}},\ }\href
  {\doibase 10.1103/PhysRevC.81.034001} {\bibfield  {journal} {\bibinfo
  {journal} {Phys. Rev. C}\ }\textbf {\bibinfo {volume} {81}},\ \bibinfo
  {pages} {034001} (\bibinfo {year} {2010})}\BibitemShut {NoStop}%
\bibitem [{\citenamefont {Yuan}\ \emph {et~al.}(2010)\citenamefont {Yuan},
  \citenamefont {Efros}, \citenamefont {Leidemann},\ and\ \citenamefont
  {Tomusiak}}]{Yuan:2010gh}%
  \BibitemOpen
  \bibfield  {author} {\bibinfo {author} {\bibfnamefont {L.~P.}\ \bibnamefont
  {Yuan}}, \bibinfo {author} {\bibfnamefont {V.~D.}\ \bibnamefont {Efros}},
  \bibinfo {author} {\bibfnamefont {W.}~\bibnamefont {Leidemann}}, \ and\
  \bibinfo {author} {\bibfnamefont {E.~L.}\ \bibnamefont {Tomusiak}},\ }\href
  {\doibase 10.1103/PhysRevC.82.054003} {\bibfield  {journal} {\bibinfo
  {journal} {Phys. Rev.}\ }\textbf {\bibinfo {volume} {C82}},\ \bibinfo {pages}
  {054003} (\bibinfo {year} {2010})},\ \Eprint {http://arxiv.org/abs/1006.3499}
  {arXiv:1006.3499 [nucl-th]} \BibitemShut {NoStop}%
\bibitem [{\citenamefont {Efros}\ \emph {et~al.}(2011)\citenamefont {Efros},
  \citenamefont {Leidemann}, \citenamefont {Orlandini},\ and\ \citenamefont
  {Tomusiak}}]{Efros_LOT:2011}%
  \BibitemOpen
  \bibfield  {author} {\bibinfo {author} {\bibfnamefont {V.~D.}\ \bibnamefont
  {Efros}}, \bibinfo {author} {\bibfnamefont {W.}~\bibnamefont {Leidemann}},
  \bibinfo {author} {\bibfnamefont {G.}~\bibnamefont {Orlandini}}, \ and\
  \bibinfo {author} {\bibfnamefont {E.~L.}\ \bibnamefont {Tomusiak}},\ }\href
  {\doibase 10.1103/PhysRevC.83.057001} {\bibfield  {journal} {\bibinfo
  {journal} {Phys. Rev. C}\ }\textbf {\bibinfo {volume} {83}},\ \bibinfo
  {pages} {057001} (\bibinfo {year} {2011})}\BibitemShut {NoStop}%
\bibitem [{\citenamefont {Yuan}\ \emph {et~al.}(2011)\citenamefont {Yuan},
  \citenamefont {Leidemann}, \citenamefont {Efros}, \citenamefont {Orlandini},\
  and\ \citenamefont {Tomusiak}}]{Yuan:2011rd}%
  \BibitemOpen
  \bibfield  {author} {\bibinfo {author} {\bibfnamefont {L.}~\bibnamefont
  {Yuan}}, \bibinfo {author} {\bibfnamefont {W.}~\bibnamefont {Leidemann}},
  \bibinfo {author} {\bibfnamefont {V.~D.}\ \bibnamefont {Efros}}, \bibinfo
  {author} {\bibfnamefont {G.}~\bibnamefont {Orlandini}}, \ and\ \bibinfo
  {author} {\bibfnamefont {E.~L.}\ \bibnamefont {Tomusiak}},\ }\href {\doibase
  10.1016/j.physletb.2011.10.066} {\bibfield  {journal} {\bibinfo  {journal}
  {Phys. Lett.}\ }\textbf {\bibinfo {volume} {B706}},\ \bibinfo {pages} {90}
  (\bibinfo {year} {2011})},\ \Eprint {http://arxiv.org/abs/1108.3204}
  {arXiv:1108.3204 [nucl-th]} \BibitemShut {NoStop}%
\bibitem [{\citenamefont {Lovato}\ \emph {et~al.}(2017)\citenamefont {Lovato},
  \citenamefont {Gandolfi}, \citenamefont {Carlson}, \citenamefont {Lusk},
  \citenamefont {Pieper},\ and\ \citenamefont {Schiavilla}}]{Lovato:2017cux}%
  \BibitemOpen
  \bibfield  {author} {\bibinfo {author} {\bibfnamefont {A.}~\bibnamefont
  {Lovato}}, \bibinfo {author} {\bibfnamefont {S.}~\bibnamefont {Gandolfi}},
  \bibinfo {author} {\bibfnamefont {J.}~\bibnamefont {Carlson}}, \bibinfo
  {author} {\bibfnamefont {E.}~\bibnamefont {Lusk}}, \bibinfo {author}
  {\bibfnamefont {S.~C.}\ \bibnamefont {Pieper}}, \ and\ \bibinfo {author}
  {\bibfnamefont {R.}~\bibnamefont {Schiavilla}},\ }\href@noop {} {\  (\bibinfo
  {year} {2017})},\ \Eprint {http://arxiv.org/abs/1711.02047} {arXiv:1711.02047
  [nucl-th]} \BibitemShut {NoStop}%
\bibitem [{\citenamefont {Rocco}\ \emph {et~al.}(2017)\citenamefont {Rocco},
  \citenamefont {Alvarez-Ruso}, \citenamefont {Lovato},\ and\ \citenamefont
  {Nieves}}]{Rocco:2017hmh}%
  \BibitemOpen
  \bibfield  {author} {\bibinfo {author} {\bibfnamefont {N.}~\bibnamefont
  {Rocco}}, \bibinfo {author} {\bibfnamefont {L.}~\bibnamefont {Alvarez-Ruso}},
  \bibinfo {author} {\bibfnamefont {A.}~\bibnamefont {Lovato}}, \ and\ \bibinfo
  {author} {\bibfnamefont {J.}~\bibnamefont {Nieves}},\ }\href {\doibase
  10.1103/PhysRevC.96.015504} {\bibfield  {journal} {\bibinfo  {journal} {Phys.
  Rev.}\ }\textbf {\bibinfo {volume} {C96}},\ \bibinfo {pages} {015504}
  (\bibinfo {year} {2017})},\ \Eprint {http://arxiv.org/abs/1701.05151}
  {arXiv:1701.05151 [nucl-th]} \BibitemShut {NoStop}%
\bibitem [{\citenamefont {Efros}\ \emph {et~al.}(1994)\citenamefont {Efros},
  \citenamefont {Leidemann},\ and\ \citenamefont {Orlandini}}]{Efros_LO:1994}%
  \BibitemOpen
  \bibfield  {author} {\bibinfo {author} {\bibfnamefont {V.~D.}\ \bibnamefont
  {Efros}}, \bibinfo {author} {\bibfnamefont {W.}~\bibnamefont {Leidemann}}, \
  and\ \bibinfo {author} {\bibfnamefont {G.}~\bibnamefont {Orlandini}},\ }\href
  {\doibase http://dx.doi.org/10.1016/0370-2693(94)91355-2} {\bibfield
  {journal} {\bibinfo  {journal} {Physics Letters B}\ }\textbf {\bibinfo
  {volume} {338}},\ \bibinfo {pages} {130 } (\bibinfo {year}
  {1994})}\BibitemShut {NoStop}%
\bibitem [{\citenamefont {Efros}\ \emph {et~al.}(2007)\citenamefont {Efros},
  \citenamefont {Leidemann}, \citenamefont {Orlandini},\ and\ \citenamefont
  {Barnea}}]{Efros:2007nq}%
  \BibitemOpen
  \bibfield  {author} {\bibinfo {author} {\bibfnamefont {V.~D.}\ \bibnamefont
  {Efros}}, \bibinfo {author} {\bibfnamefont {W.}~\bibnamefont {Leidemann}},
  \bibinfo {author} {\bibfnamefont {G.}~\bibnamefont {Orlandini}}, \ and\
  \bibinfo {author} {\bibfnamefont {N.}~\bibnamefont {Barnea}},\ }\href
  {\doibase 10.1088/0954-3899/34/12/R02} {\bibfield  {journal} {\bibinfo
  {journal} {J. Phys.}\ }\textbf {\bibinfo {volume} {G34}},\ \bibinfo {pages}
  {R459} (\bibinfo {year} {2007})},\ \Eprint {http://arxiv.org/abs/0708.2803}
  {arXiv:0708.2803 [nucl-th]} \BibitemShut {NoStop}%
\bibitem [{\citenamefont {Shen}\ \emph {et~al.}(2012)\citenamefont {Shen},
  \citenamefont {Marcucci}, \citenamefont {Carlson}, \citenamefont {Gandolfi},\
  and\ \citenamefont {Schiavilla}}]{Shen:2012}%
  \BibitemOpen
  \bibfield  {author} {\bibinfo {author} {\bibfnamefont {G.}~\bibnamefont
  {Shen}}, \bibinfo {author} {\bibfnamefont {L.}~\bibnamefont {Marcucci}},
  \bibinfo {author} {\bibfnamefont {J.}~\bibnamefont {Carlson}}, \bibinfo
  {author} {\bibfnamefont {S.}~\bibnamefont {Gandolfi}}, \ and\ \bibinfo
  {author} {\bibfnamefont {R.}~\bibnamefont {Schiavilla}},\ }\href {\doibase
  10.1103/PhysRevC.86.035503} {\bibfield  {journal} {\bibinfo  {journal} {Phys.
  Rev. C}\ }\textbf {\bibinfo {volume} {86}},\ \bibinfo {pages} {035503}
  (\bibinfo {year} {2012})}\BibitemShut {NoStop}%
\bibitem [{\citenamefont {Carlson}\ \emph {et~al.}(2002)\citenamefont
  {Carlson}, \citenamefont {Jourdan}, \citenamefont {Schiavilla},\ and\
  \citenamefont {Sick}}]{Carlson:2002}%
  \BibitemOpen
  \bibfield  {author} {\bibinfo {author} {\bibfnamefont {J.}~\bibnamefont
  {Carlson}}, \bibinfo {author} {\bibfnamefont {J.}~\bibnamefont {Jourdan}},
  \bibinfo {author} {\bibfnamefont {R.}~\bibnamefont {Schiavilla}}, \ and\
  \bibinfo {author} {\bibfnamefont {I.}~\bibnamefont {Sick}},\ }\href {\doibase
  10.1103/PhysRevC.65.024002} {\bibfield  {journal} {\bibinfo  {journal} {Phys.
  Rev. C}\ }\textbf {\bibinfo {volume} {65}},\ \bibinfo {pages} {024002}
  (\bibinfo {year} {2002})}\BibitemShut {NoStop}%
\bibitem [{\citenamefont {Wiringa}\ \emph {et~al.}(1995)\citenamefont
  {Wiringa}, \citenamefont {Stoks},\ and\ \citenamefont
  {Schiavilla}}]{Wiringa:1995}%
  \BibitemOpen
  \bibfield  {author} {\bibinfo {author} {\bibfnamefont {R.~B.}\ \bibnamefont
  {Wiringa}}, \bibinfo {author} {\bibfnamefont {V.~G.~J.}\ \bibnamefont
  {Stoks}}, \ and\ \bibinfo {author} {\bibfnamefont {R.}~\bibnamefont
  {Schiavilla}},\ }\href {\doibase 10.1103/PhysRevC.51.38} {\bibfield
  {journal} {\bibinfo  {journal} {Phys. Rev. C}\ }\textbf {\bibinfo {volume}
  {51}},\ \bibinfo {pages} {38} (\bibinfo {year} {1995})}\BibitemShut {NoStop}%
\bibitem [{\citenamefont {Pieper}\ and\ \citenamefont
  {Wiringa}(2001)}]{Pieper:2001}%
  \BibitemOpen
  \bibfield  {author} {\bibinfo {author} {\bibfnamefont {S.~C.}\ \bibnamefont
  {Pieper}}\ and\ \bibinfo {author} {\bibfnamefont {R.~B.}\ \bibnamefont
  {Wiringa}},\ }\href {\doibase 10.1146/annurev.nucl.51.101701.132506}
  {\bibfield  {journal} {\bibinfo  {journal} {Ann.Rev.Nucl.Part.Sci.}\ }\textbf
  {\bibinfo {volume} {51}},\ \bibinfo {pages} {53} (\bibinfo {year} {2001})},\
  \Eprint {http://arxiv.org/abs/nucl-th/0103005} {arXiv:nucl-th/0103005
  [nucl-th]} \BibitemShut {NoStop}%
\bibitem [{\citenamefont {Simon}\ \emph {et~al.}(1980)\citenamefont {Simon},
  \citenamefont {Schmitt}, \citenamefont {Borkowski},\ and\ \citenamefont
  {Walther}}]{Simon:1980hu}%
  \BibitemOpen
  \bibfield  {author} {\bibinfo {author} {\bibfnamefont {G.~G.}\ \bibnamefont
  {Simon}}, \bibinfo {author} {\bibfnamefont {C.}~\bibnamefont {Schmitt}},
  \bibinfo {author} {\bibfnamefont {F.}~\bibnamefont {Borkowski}}, \ and\
  \bibinfo {author} {\bibfnamefont {V.~H.}\ \bibnamefont {Walther}},\ }\href
  {\doibase 10.1016/0375-9474(80)90104-9} {\bibfield  {journal} {\bibinfo
  {journal} {Nucl. Phys.}\ }\textbf {\bibinfo {volume} {A333}},\ \bibinfo
  {pages} {381} (\bibinfo {year} {1980})}\BibitemShut {NoStop}%
\bibitem [{\citenamefont {Galster}\ \emph {et~al.}(1971)\citenamefont
  {Galster}, \citenamefont {Klein}, \citenamefont {Moritz}, \citenamefont
  {Schmidt}, \citenamefont {Wegener},\ and\ \citenamefont
  {Bleckwenn}}]{Galster:1971kv}%
  \BibitemOpen
  \bibfield  {author} {\bibinfo {author} {\bibfnamefont {S.}~\bibnamefont
  {Galster}}, \bibinfo {author} {\bibfnamefont {H.}~\bibnamefont {Klein}},
  \bibinfo {author} {\bibfnamefont {J.}~\bibnamefont {Moritz}}, \bibinfo
  {author} {\bibfnamefont {K.~H.}\ \bibnamefont {Schmidt}}, \bibinfo {author}
  {\bibfnamefont {D.}~\bibnamefont {Wegener}}, \ and\ \bibinfo {author}
  {\bibfnamefont {J.}~\bibnamefont {Bleckwenn}},\ }\href {\doibase
  10.1016/0550-3213(71)90068-X} {\bibfield  {journal} {\bibinfo  {journal}
  {Nucl. Phys.}\ }\textbf {\bibinfo {volume} {B32}},\ \bibinfo {pages} {221}
  (\bibinfo {year} {1971})}\BibitemShut {NoStop}%
\bibitem [{\citenamefont {Hohler}\ \emph {et~al.}(1976)\citenamefont {Hohler},
  \citenamefont {Pietarinen}, \citenamefont {Sabba~Stefanescu}, \citenamefont
  {Borkowski}, \citenamefont {Simon}, \citenamefont {Walther},\ and\
  \citenamefont {Wendling}}]{Hohler:1976ax}%
  \BibitemOpen
  \bibfield  {author} {\bibinfo {author} {\bibfnamefont {G.}~\bibnamefont
  {Hohler}}, \bibinfo {author} {\bibfnamefont {E.}~\bibnamefont {Pietarinen}},
  \bibinfo {author} {\bibfnamefont {I.}~\bibnamefont {Sabba~Stefanescu}},
  \bibinfo {author} {\bibfnamefont {F.}~\bibnamefont {Borkowski}}, \bibinfo
  {author} {\bibfnamefont {G.~G.}\ \bibnamefont {Simon}}, \bibinfo {author}
  {\bibfnamefont {V.~H.}\ \bibnamefont {Walther}}, \ and\ \bibinfo {author}
  {\bibfnamefont {R.~D.}\ \bibnamefont {Wendling}},\ }\href {\doibase
  10.1016/0550-3213(76)90449-1} {\bibfield  {journal} {\bibinfo  {journal}
  {Nucl. Phys.}\ }\textbf {\bibinfo {volume} {B114}},\ \bibinfo {pages} {505}
  (\bibinfo {year} {1976})}\BibitemShut {NoStop}%
\bibitem [{\citenamefont {Lovato}\ \emph {et~al.}(2013)\citenamefont {Lovato},
  \citenamefont {Gandolfi}, \citenamefont {Butler}, \citenamefont {Carlson},
  \citenamefont {Lusk}, \citenamefont {Pieper},\ and\ \citenamefont
  {Schiavilla}}]{Lovato:2013}%
  \BibitemOpen
  \bibfield  {author} {\bibinfo {author} {\bibfnamefont {A.}~\bibnamefont
  {Lovato}}, \bibinfo {author} {\bibfnamefont {S.}~\bibnamefont {Gandolfi}},
  \bibinfo {author} {\bibfnamefont {R.}~\bibnamefont {Butler}}, \bibinfo
  {author} {\bibfnamefont {J.}~\bibnamefont {Carlson}}, \bibinfo {author}
  {\bibfnamefont {E.}~\bibnamefont {Lusk}}, \bibinfo {author} {\bibfnamefont
  {S.~C.}\ \bibnamefont {Pieper}}, \ and\ \bibinfo {author} {\bibfnamefont
  {R.}~\bibnamefont {Schiavilla}},\ }\href {\doibase
  10.1103/PhysRevLett.111.092501} {\bibfield  {journal} {\bibinfo  {journal}
  {Phys. Rev. Lett.}\ }\textbf {\bibinfo {volume} {111}},\ \bibinfo {pages}
  {092501} (\bibinfo {year} {2013})}\BibitemShut {NoStop}%
\bibitem [{\citenamefont {Carlson}\ \emph
  {et~al.}(2015{\natexlab{b}})\citenamefont {Carlson}, \citenamefont
  {Gandolfi}, \citenamefont {Pederiva}, \citenamefont {Pieper}, \citenamefont
  {Schiavilla}, \citenamefont {Schmidt},\ and\ \citenamefont
  {Wiringa}}]{Carlson:2015}%
  \BibitemOpen
  \bibfield  {author} {\bibinfo {author} {\bibfnamefont {J.}~\bibnamefont
  {Carlson}}, \bibinfo {author} {\bibfnamefont {S.}~\bibnamefont {Gandolfi}},
  \bibinfo {author} {\bibfnamefont {F.}~\bibnamefont {Pederiva}}, \bibinfo
  {author} {\bibfnamefont {S.~C.}\ \bibnamefont {Pieper}}, \bibinfo {author}
  {\bibfnamefont {R.}~\bibnamefont {Schiavilla}}, \bibinfo {author}
  {\bibfnamefont {K.~E.}\ \bibnamefont {Schmidt}}, \ and\ \bibinfo {author}
  {\bibfnamefont {R.~B.}\ \bibnamefont {Wiringa}},\ }\href {\doibase
  10.1103/RevModPhys.87.1067} {\bibfield  {journal} {\bibinfo  {journal} {Rev.
  Mod. Phys.}\ }\textbf {\bibinfo {volume} {87}},\ \bibinfo {pages} {1067}
  (\bibinfo {year} {2015}{\natexlab{b}})}\BibitemShut {NoStop}%
\bibitem [{\citenamefont {Bacca}\ \emph
  {et~al.}(2009{\natexlab{a}})\citenamefont {Bacca}, \citenamefont {Barnea},
  \citenamefont {Leidemann},\ and\ \citenamefont {Orlandini}}]{Bacca:2009}%
  \BibitemOpen
  \bibfield  {author} {\bibinfo {author} {\bibfnamefont {S.}~\bibnamefont
  {Bacca}}, \bibinfo {author} {\bibfnamefont {N.}~\bibnamefont {Barnea}},
  \bibinfo {author} {\bibfnamefont {W.}~\bibnamefont {Leidemann}}, \ and\
  \bibinfo {author} {\bibfnamefont {G.}~\bibnamefont {Orlandini}},\ }\href
  {\doibase 10.1103/PhysRevLett.102.162501} {\bibfield  {journal} {\bibinfo
  {journal} {Phys. Rev. Lett.}\ }\textbf {\bibinfo {volume} {102}},\ \bibinfo
  {pages} {162501} (\bibinfo {year} {2009}{\natexlab{a}})}\BibitemShut
  {NoStop}%
\bibitem [{\citenamefont {Bacca}\ \emph
  {et~al.}(2009{\natexlab{b}})\citenamefont {Bacca}, \citenamefont {Barnea},
  \citenamefont {Leidemann},\ and\ \citenamefont {Orlandini}}]{Bacca:2009prc}%
  \BibitemOpen
  \bibfield  {author} {\bibinfo {author} {\bibfnamefont {S.}~\bibnamefont
  {Bacca}}, \bibinfo {author} {\bibfnamefont {N.}~\bibnamefont {Barnea}},
  \bibinfo {author} {\bibfnamefont {W.}~\bibnamefont {Leidemann}}, \ and\
  \bibinfo {author} {\bibfnamefont {G.}~\bibnamefont {Orlandini}},\ }\href
  {\doibase 10.1103/PhysRevC.80.064001} {\bibfield  {journal} {\bibinfo
  {journal} {Phys. Rev. C}\ }\textbf {\bibinfo {volume} {80}},\ \bibinfo
  {pages} {064001} (\bibinfo {year} {2009}{\natexlab{b}})}\BibitemShut
  {NoStop}%
\bibitem [{\citenamefont {Bacca}\ \emph {et~al.}(2013)\citenamefont {Bacca},
  \citenamefont {Barnea}, \citenamefont {Leidemann},\ and\ \citenamefont
  {Orlandini}}]{Bacca:2012xv}%
  \BibitemOpen
  \bibfield  {author} {\bibinfo {author} {\bibfnamefont {S.}~\bibnamefont
  {Bacca}}, \bibinfo {author} {\bibfnamefont {N.}~\bibnamefont {Barnea}},
  \bibinfo {author} {\bibfnamefont {W.}~\bibnamefont {Leidemann}}, \ and\
  \bibinfo {author} {\bibfnamefont {G.}~\bibnamefont {Orlandini}},\ }\href
  {\doibase 10.1103/PhysRevLett.110.042503} {\bibfield  {journal} {\bibinfo
  {journal} {Phys. Rev. Lett.}\ }\textbf {\bibinfo {volume} {110}},\ \bibinfo
  {pages} {042503} (\bibinfo {year} {2013})},\ \Eprint
  {http://arxiv.org/abs/1210.7255} {arXiv:1210.7255 [nucl-th]} \BibitemShut
  {NoStop}%
\bibitem [{\citenamefont {Barnea}\ \emph {et~al.}(2010)\citenamefont {Barnea},
  \citenamefont {Efros}, \citenamefont {Leidemann},\ and\ \citenamefont
  {Orlandini}}]{Barnea:2009zu}%
  \BibitemOpen
  \bibfield  {author} {\bibinfo {author} {\bibfnamefont {N.}~\bibnamefont
  {Barnea}}, \bibinfo {author} {\bibfnamefont {V.~D.}\ \bibnamefont {Efros}},
  \bibinfo {author} {\bibfnamefont {W.}~\bibnamefont {Leidemann}}, \ and\
  \bibinfo {author} {\bibfnamefont {G.}~\bibnamefont {Orlandini}},\ }\href
  {\doibase 10.1007/s00601-009-0081-0} {\bibfield  {journal} {\bibinfo
  {journal} {Few Body Syst.}\ }\textbf {\bibinfo {volume} {47}},\ \bibinfo
  {pages} {201} (\bibinfo {year} {2010})},\ \Eprint
  {http://arxiv.org/abs/0906.5421} {arXiv:0906.5421 [nucl-th]} \BibitemShut
  {NoStop}%
\bibitem [{\citenamefont {Ritz}\ \emph {et~al.}(1997)\citenamefont {Ritz},
  \citenamefont {Goller}, \citenamefont {Wilbois},\ and\ \citenamefont
  {Arenhovel}}]{Ritz:1996za}%
  \BibitemOpen
  \bibfield  {author} {\bibinfo {author} {\bibfnamefont {F.}~\bibnamefont
  {Ritz}}, \bibinfo {author} {\bibfnamefont {H.}~\bibnamefont {Goller}},
  \bibinfo {author} {\bibfnamefont {T.}~\bibnamefont {Wilbois}}, \ and\
  \bibinfo {author} {\bibfnamefont {H.}~\bibnamefont {Arenhovel}},\ }\href
  {\doibase 10.1103/PhysRevC.55.2214} {\bibfield  {journal} {\bibinfo
  {journal} {Phys. Rev.}\ }\textbf {\bibinfo {volume} {C55}},\ \bibinfo {pages}
  {2214} (\bibinfo {year} {1997})},\ \Eprint
  {http://arxiv.org/abs/nucl-th/9611026} {arXiv:nucl-th/9611026 [nucl-th]}
  \BibitemShut {NoStop}%
\bibitem [{\citenamefont {Rocco}\ \emph {et~al.}(2016)\citenamefont {Rocco},
  \citenamefont {Lovato},\ and\ \citenamefont {Benhar}}]{Rocco:2016ejr}%
  \BibitemOpen
  \bibfield  {author} {\bibinfo {author} {\bibfnamefont {N.}~\bibnamefont
  {Rocco}}, \bibinfo {author} {\bibfnamefont {A.}~\bibnamefont {Lovato}}, \
  and\ \bibinfo {author} {\bibfnamefont {O.}~\bibnamefont {Benhar}},\ }\href
  {\doibase 10.1103/PhysRevC.94.065501} {\bibfield  {journal} {\bibinfo
  {journal} {Phys. Rev.}\ }\textbf {\bibinfo {volume} {C94}},\ \bibinfo {pages}
  {065501} (\bibinfo {year} {2016})},\ \Eprint
  {http://arxiv.org/abs/1610.06081} {arXiv:1610.06081 [nucl-th]} \BibitemShut
  {NoStop}%
\bibitem [{\citenamefont {Alberico}\ \emph {et~al.}(1988)\citenamefont
  {Alberico}, \citenamefont {Molinari}, \citenamefont {Donnelly}, \citenamefont
  {Kronenberg},\ and\ \citenamefont {Van~Orden}}]{Alberico:1988bv}%
  \BibitemOpen
  \bibfield  {author} {\bibinfo {author} {\bibfnamefont {W.~M.}\ \bibnamefont
  {Alberico}}, \bibinfo {author} {\bibfnamefont {A.}~\bibnamefont {Molinari}},
  \bibinfo {author} {\bibfnamefont {T.~W.}\ \bibnamefont {Donnelly}}, \bibinfo
  {author} {\bibfnamefont {E.~L.}\ \bibnamefont {Kronenberg}}, \ and\ \bibinfo
  {author} {\bibfnamefont {J.~W.}\ \bibnamefont {Van~Orden}},\ }\href {\doibase
  10.1103/PhysRevC.38.1801} {\bibfield  {journal} {\bibinfo  {journal} {Phys.
  Rev.}\ }\textbf {\bibinfo {volume} {C38}},\ \bibinfo {pages} {1801} (\bibinfo
  {year} {1988})}\BibitemShut {NoStop}%
\bibitem [{\citenamefont {Barbaro}\ \emph {et~al.}(1998)\citenamefont
  {Barbaro}, \citenamefont {Cenni}, \citenamefont {De~Pace}, \citenamefont
  {Donnelly},\ and\ \citenamefont {Molinari}}]{Barbaro:1998gu}%
  \BibitemOpen
  \bibfield  {author} {\bibinfo {author} {\bibfnamefont {M.~B.}\ \bibnamefont
  {Barbaro}}, \bibinfo {author} {\bibfnamefont {R.}~\bibnamefont {Cenni}},
  \bibinfo {author} {\bibfnamefont {A.}~\bibnamefont {De~Pace}}, \bibinfo
  {author} {\bibfnamefont {T.~W.}\ \bibnamefont {Donnelly}}, \ and\ \bibinfo
  {author} {\bibfnamefont {A.}~\bibnamefont {Molinari}},\ }\href {\doibase
  10.1016/S0375-9474(98)00443-6} {\bibfield  {journal} {\bibinfo  {journal}
  {Nucl. Phys.}\ }\textbf {\bibinfo {volume} {A643}},\ \bibinfo {pages} {137}
  (\bibinfo {year} {1998})},\ \Eprint {http://arxiv.org/abs/nucl-th/9804054}
  {arXiv:nucl-th/9804054 [nucl-th]} \BibitemShut {NoStop}%
\bibitem [{\citenamefont {Benhar}\ \emph {et~al.}(2006)\citenamefont {Benhar},
  \citenamefont {Day},\ and\ \citenamefont {Sick}}]{Benhar:2006er}%
  \BibitemOpen
  \bibfield  {author} {\bibinfo {author} {\bibfnamefont {O.}~\bibnamefont
  {Benhar}}, \bibinfo {author} {\bibfnamefont {D.}~\bibnamefont {Day}}, \ and\
  \bibinfo {author} {\bibfnamefont {I.}~\bibnamefont {Sick}},\ }\href@noop {}
  {\  (\bibinfo {year} {2006})},\ \Eprint
  {http://arxiv.org/abs/nucl-ex/0603032} {arXiv:nucl-ex/0603032 [nucl-ex]}
  \BibitemShut {NoStop}%
\end{thebibliography}%
\end{document}